\documentclass[aps,prx,superscriptaddress,twocolumn,longbibliography,floatfix]{revtex4-2}
\usepackage{units}
\usepackage{amsmath}
\usepackage{amsthm}
\usepackage{amssymb}
\usepackage{graphicx}
\usepackage{color}
\usepackage{xcolor}
\usepackage{bbold}

\definecolor{myurlcolor}{rgb}{0,0,0.7}
\definecolor{myrefcolor}{rgb}{0.8,0,0}

\usepackage[
	breaklinks,
	pdftex,
	colorlinks=true, 
	linkcolor=myrefcolor,
	citecolor=myurlcolor,
	urlcolor=myurlcolor
]{hyperref}

\usepackage[linesnumbered, ruled,vlined]{algorithm2e}

\graphicspath{{./images/},{./imagesAppendix/}}

\renewcommand{\eqref}[1]{Eq.~(\ref{#1})} 

\def\app#1#2{%
  \mathrel{%
    \setbox0=\hbox{$#1\sim$}%
    \setbox2=\hbox{%
      \rlap{\hbox{$#1\propto$}}%
      \lower1.1\ht0\box0%
    }%
    \raise0.25\ht2\box2%
  }%
}
\def\approxprop{\mathpalette\app\relax}

\theoremstyle{plain}

\theoremstyle{plain}

\ifx\proof\undefined
\newenvironment{proof}[1][\protect\proofname]{\par
	\normalfont\topsep6\p@\@plus6\p@\relax
	\trivlist
	\itemindent\parindent
	\item[\hskip\labelsep\scshape #1]\ignorespaces
}{%
	\endtrivlist\@endpefalse
}
\providecommand{\proofname}{Proof}
\fi
\theoremstyle{plain}

\theoremstyle{remark}

\newcommand{\bra}[1]{\langle #1|}
\newcommand{\ket}[1]{|#1 \rangle}
\makeatother
\newcommand{\braket}[2]{\langle #1 \vert #2 \rangle}
\newcommand{\abs}[1]{\left|#1\right|}
\newcommand{\idg}[1]{{\bfseries #1)}}

\newcommand\numberthis{\addtocounter{equation}{1}\tag{\theequation}}
\providecommand{\factname}{Fact}
\providecommand{\theoremname}{Theorem}
\providecommand{\claimname}{Claim}
\providecommand{\lemmaname}{Lemma}
\providecommand{\definitionname}{Definition}

\definecolor{KB}{rgb}{0.4,0.3,0.9}

\definecolor{THc}{rgb}{0.9,0.3,0.2}

\newcommand{\revA}[1]{{#1}}
\newcommand{\revB}[1]{{#1}}

\theoremstyle{definition}

\newcommand{\subfigimg}[3][,]{%
	\setbox1=\hbox{\includegraphics[#1]{#3}}
	\leavevmode\rlap{\usebox1}
	\rlap{\hspace*{2pt}\raisebox{\dimexpr\ht1-0.5\baselineskip}{{\bfseries \large\textsf{#2}}}}
	\phantom{\usebox1}
}

\newcommand{\sectionMain}[1]{
\let\oldaddcontentsline\addcontentsline
\renewcommand{\addcontentsline}[3]{}
\section{#1}
\let\addcontentsline\oldaddcontentsline
}

\allowdisplaybreaks

\begin{document}

\title{Scalable measures of magic resource for quantum computers}

\author{Tobias Haug}

\email{thaug@ic.ac.uk}
\affiliation{QOLS, Blackett Laboratory, Imperial College London SW7 2AZ, UK}

\author{M. S. Kim}
\affiliation{QOLS, Blackett Laboratory, Imperial College London SW7 2AZ, UK}
\begin{abstract}
Non-stabilizerness or magic resource characterizes the amount of non-Clifford operations needed to prepare quantum states.
It is a crucial resource for quantum computing and a necessary condition for quantum advantage. 
However, quantifying magic resource beyond a few qubits has been a major challenge.
Here, we introduce efficient measures of magic resource for pure quantum states with a sampling cost that is independent of the number of qubits.
Our method uses Bell measurements over two copies of a state, which we implement in experiment together with a cost-free error mitigation scheme.
We show the transition of classically simulable stabilizer states into intractable quantum states on the IonQ quantum computer.
For applications, we efficiently distinguish stabilizer and non-stabilizer states with low measurement cost even in the presence of experimental noise. Further, we propose a variational quantum algorithm to maximize our measure via the shift-rule. Our algorithm can be free of barren plateaus even for highly expressible variational circuits. Finally, we experimentally demonstrate a Bell measurement protocol for the stabilizer Rényi entropy as well as the Wallach-Meyer entanglement measure.
Our results pave the way to understand the non-classical power of quantum computers, quantum simulators and quantum many-body systems.
\end{abstract}

\maketitle

\sectionMain{Introduction}
Simulating quantum states is in general intractable for classical computers. However, particular classes of states can be efficiently simulated classically. \revA{An important example are stabilizer states which are generated from Clifford operations~\cite{gottesman1998heisenberg,aaronson2004improved}. The number of non-Clifford operations needed to prepare a state can be quantified by measures of non-stabilizerness or magic resource.} \revB{Henceforth we will abbreviate the term ``magic resource'' with ``magic''.}
Magic can be related to the difficulty of classical simulation of quantum states~\cite{howard2014contextuality,pashayan2015estimating,bravyi2016trading,bravyi2019simulation,seddon2021quantifying,seddon2019quantifying2,koukoulekidis2022born} and to quantum chaos~\cite{leone2021renyi,leone2021quantum,haferkamp2022quantum}. Further, magic is a precious resource required to realize universal unitaries~\cite{bravyi2005universal,campbell2017roads} in fault-tolerant quantum computers~\cite{shor1996fault,preskill1998fault,gottesman1999demonstrating,kitaev2003fault}.

To characterize magic, various measures have been proposed~\cite{campbell2011catalysis,howard2017application,hahn2021quantifying,bravyi2019simulation,veitch2014resource,heinrich2019robustness,wang2019quantifying,beverland2020lower,leone2021renyi,saxena2022quantifying,dai2022detecting,delfosse2015wigner}. However, most measures require solving an optimization program, access to the amplitudes of the quantum state and a computational cost that scales exponentially with the system size. Recently, stabilizer entropy was proposed as an experimentally accessible measure~\cite{leone2021renyi}, however its randomized measurement protocol scales exponentially with the number of qubits~\cite{brydges2019probing}. 

Rapid progress has been made in experimental demonstrations of noisy intermediate-scale quantum computers~\cite{preskill2018quantum,bharti2021noisy} and fault-tolerant quantum computers~\cite{ryan2021realization,egan2021fault,zhao2021realizing,krinner2021realizing}. A major challenge is to benchmark the power of quantum computers and track their progress~\cite{eisert2020quantum,carrasco2021theoretical}.
For fault-tolerant quantum computers, magic characterizes the capability to implement universal quantum gates~\cite{campbell2017roads}.
For noisy intermediate-scale quantum computers, an important benchmark is to prepare states that are difficult to simulate classically~\cite{arute2019quantum} which can related to particular measures of magic~\cite{pashayan2015estimating,bravyi2016trading,bravyi2019simulation,leone2021quantum}.  The relationship between complexity of quantum states and magic is also of major interest in quantum many-body physics~\cite{liu2020many,white2021conformal}.

Here, we introduce Bell magic as an efficiently computable measure of magic for quantum computers. The number of measurements is independent of the number of qubits, and the classical post-processing time scales linearly. 
\revB{Bell magic is a faithful measure of magic for pure states, while for mixed states it is not faithful in general. For noisy quantum computers, we use Bell measurements over two copies of a state and error mitigation to compute Bell magic efficiently.}
We propose practical applications of Bell magic for state discrimination and finding highly magical states with variational quantum algorithms. Our variational quantum algorithm can have large gradients even for highly expressible ansatz circuits.
With the IonQ quantum computer, we study the transition from stabilizer states to intractable quantum states. Further, we experimentally distinguish different types of states using magic.
Our results provide an indispensable tool to characterize the magic of quantum computers, quantum simulators and numerical simulations of quantum many-body systems.

We first define preliminary concepts in Sec.\ref{sec:prelim}. Then, we introduce Bell magic in Sec.\ref{sec:bellmagic}, its measurement scheme in Sec.\ref{sec:measuremagic} and the method to mitigate errors in Sec.\ref{sec:errormitigation}. We numerically and experimentally demonstrate the measurement of Bell magic in Sec.\ref{sec:results}. Then, we show applications of Bell magic for state discrimination in Sec.\ref{sec:discrimination} and finding highly magical states with variational quantum algorithms in Sec.\ref{sec:magicvariational}. Finally, the results are discussed in Sec.\ref{sec:discussion}.
We give an overview of the definitions of symbols in Tab.\ref{tab:definitions}.

\sectionMain{Preliminaries}\label{sec:prelim}

\begin{figure*}[htbp]
	\centering	
	\subfigimg[width=0.97\textwidth]{}{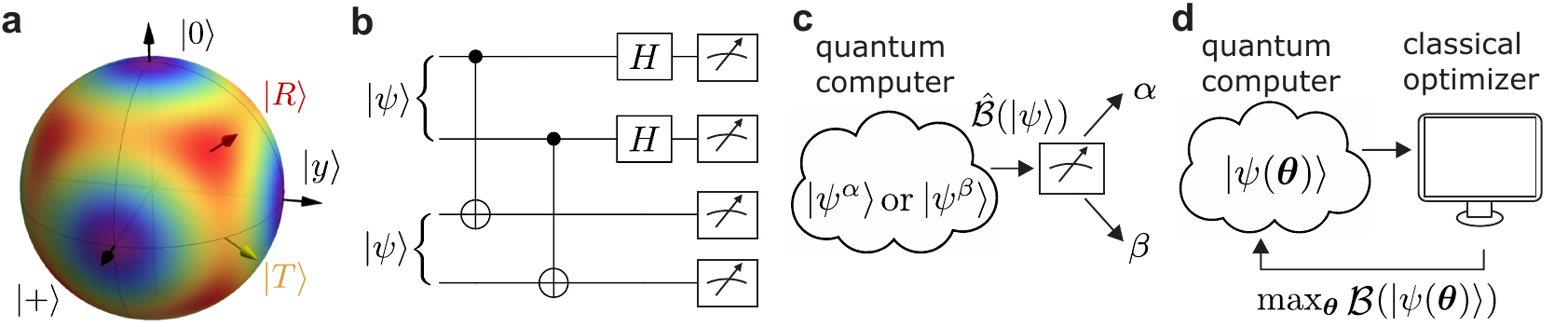} 
	\caption{\idg{a} 
    Bell magic $\mathcal{B}$ (\eqref{eq:BellMagic}) is a measure of non-stabilizerness for pure quantum states. Color shows $\mathcal{B}$ for the Bloch sphere of a single qubit with~\eqref{eq:productBell}, where magnitude of $\mathcal{B}$ is increasing from blue to red. Stabilizer states along the main axes such as $\ket{0}$, $\ket{+}$ and $\ket{y}$ have zero magic, while the magic states $\ket{T}$ and $\ket{R}$ have non-zero magic. \idg{b} Quantum circuit to measure $\mathcal{B}$ of $N$-qubit state $\ket{\psi}=U\ket{0}$ with unitary $U$ by preparing two copies $\ket{\psi}\otimes\ket{\psi}$ and measuring in the Bell basis. The number of required measurements $N_\text{Q}$ is independent of $N$.
	\idg{c} State discrimination scheme to determine whether a given state $\ket{\psi}$ belongs to one of two classes $\alpha,\beta$, with respective magic $\mathcal{B}^\alpha>\mathcal{B}^\beta$. When the estimated Bell magic $\hat{\mathcal{B}}$ is greater than a threshold $\hat{\mathcal{B}}>\mathcal{B}^*$, we classify the state as class $\alpha$, else class $\beta$.
	\idg{d} Variational quantum algorithm to maximize magic by optimizing parameter $\boldsymbol{\theta}$ of parameterized quantum circuit $\ket{\psi(\boldsymbol{\theta})}=U(\boldsymbol{\theta})\ket{0}$. The algorithm runs via feed-back loop between measurements on the quantum computer and a classical optimization routine. 
	}
	\label{fig:sketch}
\end{figure*}

We define the Pauli matrices $\sigma_{00}=I_{2}$, $\sigma_{01}=\sigma^x$, $\sigma_{10}=\sigma^z$ and $\sigma_{11}=\sigma^y$.
The $4^N$ Pauli strings are $N$-qubit tensor products of Pauli matrices which we define as $\sigma_{\boldsymbol{n}}=\bigotimes_{j=1}^N \sigma_{\boldsymbol{n}_{2j-1}\boldsymbol{n}_{2j}}$ with $\boldsymbol{n}\in\{0,1\}^{2N}$.
The product of two Pauli strings $\sigma_{\boldsymbol{r}}$, $\sigma_{\boldsymbol{q}}$ can be written as $\sigma_{\boldsymbol{r}}\sigma_{\boldsymbol{q}}=\sigma_{\boldsymbol{r}\oplus\boldsymbol{q}}$ up to a multiplication with $\{\pm 1,\pm i\}$, where $\oplus$ denotes a bit-wise exclusive or.
The Bell states are given by $\ket{\sigma_{00}} = \frac{1}{\sqrt{2}} \left( \ket{00} + \ket{11} \right)$, $\ket{\sigma_{01}}= \frac{1}{\sqrt{2}} \left( \ket{00} - \ket{11} \right)$, $\ket{\sigma_{10}} = \frac{1}{\sqrt{2}} \left( \ket{01} + \ket{10} \right)$ and $\ket{\sigma_{11}}=\frac{1}{\sqrt{2}} \left( \ket{01} - \ket{10} \right)$ and we define the product of Bell states $\ket{\sigma_{\boldsymbol{r}}}=\ket{\sigma_{r_1r_2}}\otimes\dots\otimes\ket{\sigma_{r_{2N-1}r_{2N}}}$. 

Stabilizer states $\ket{\psi_\text{STAB}}$ are defined by a commuting subgroup $G$ of $\abs{G}=2^N$ Pauli strings $\sigma$. We have $\bra{\psi_\text{STAB}}\sigma\ket{\psi_\text{STAB}}=\pm1$ for $\sigma\in G$ and  $\bra{\psi_\text{STAB}}\sigma'\ket{\psi_\text{STAB}}=0$ for $\sigma'\notin G$~\cite{gottesman1998heisenberg}. Any $\sigma_{\boldsymbol{r}},\sigma_{\boldsymbol{r}'}\in G$ commute $[\sigma_{\boldsymbol{r}},\sigma_{\boldsymbol{r}'}]=0$. Unitaries that transform stabilizer states into stabilizer states are the Clifford circuits $U_\text{C}$. They can be generated by combining the Clifford gate set consisting of the $S$-gate ($S=\text{diag}[1,\exp(-i\pi/2)]$), Hadamard gate and CNOT gate, which can be efficiently simulated on classical computers~\cite{gottesman1998heisenberg}. Universal unitaries are realized by combining Clifford circuits with non-Clifford resources such as the $T$-gate ($T=\text{diag}[1,\exp(-i\pi/4)]$)~\cite{nielsen2002quantum}. Examples of stabilizer and non-stabilizer single qubit states are shown in Fig.~\ref{fig:sketch}a.

Measures of magic characterize the distance to the set of stabilizers states or unitaries~\cite{campbell2011catalysis,howard2017application,hahn2021quantifying,bravyi2019simulation,veitch2014resource,heinrich2019robustness,wang2019quantifying,leone2021renyi,saxena2022quantifying,dai2022detecting}. \revA{Measures of magic are zero for stabilizer states, and greater zero else. Further, they should be non-increasing under Clifford operations~\cite{veitch2014resource}.
Most schemes for fault-tolerant quantum computers are based on stabilizers~\cite{shor1996fault}, where universal quantum computation is enabled by consuming magic states~\cite{bravyi2005universal}. Lower bounds on the number of magic states necessary to generate a state or unitary can be related to (sub-)additive measures of magic such as the robustness of magic~\cite{howard2017application}, stabilizer entropy~\cite{leone2021renyi} or mana~\cite{veitch2014resource}.
Further, measures such as the stabilizer rank~\cite{bravyi2016trading}, robustness of magic~\cite{howard2017application} or negativity~\cite{pashayan2015estimating} can be related to the computational difficulty of particular simulation algorithms for quantum states. For these algorithms, the simulation cost increases drastically with the number of non-Clifford gates.
To understand the quantum and classical cost of simulating and preparing quantum states, one would like to compute measures of magic for large quantum systems. However,  the computational cost scales in general exponentially with qubit number for aforementioned measures. }

We now introduce a measure of magic that can be efficiently computed. We make use of entangled measurements over multiple copies of states which can reveal information not accessible by single copies~\cite{ekert2002direct,aharonov2021quantum,chen2021exponential,huang2021demonstrating}.
In particular, measurements in the basis of Bell states are known to give access to important properties which for single copies would require exponential resources~\cite{garcia2013swap,harrow2013testing,islam2015measuring,montanaro2017learning,huang2021demonstrating}. To realize Bell measurements, we first prepare the tensor product $\rho_\text{A}\otimes\rho_\text{B}$ of two states $\rho_\text{A}$, $\rho_\text{B}$.
Then, we apply the Bell transformation with the unitary $U_\text{Bell}=\bigotimes_{n=1}^N (H\otimes I_2)\cdot\text{CNOT}$ on all $N$ qubit pairs (see Fig.\ref{fig:sketch}b). 
This transformation can also be realized in atomic or photonic systems with a beam splitter~\cite{islam2015measuring}. 
Then, we measure $N_\text{Q}$ times in the computational basis and record the outcomes $\boldsymbol{r}^j\in\{0,1\}^{2N}$ with $j=1,\dots,N_\text{Q}$. Here $\boldsymbol{r}^j_{2n-1}$, $n=1,\dots,N$ is the outcome of the $n$th qubit of subsystem $A$, and $\boldsymbol{r}^j_{2n}$ of subsystem $B$. 
This measurement setting realizes a SWAP test to compute the trace overlap of the two states
\begin{equation}\label{eq:SWAP}
\text{tr}(\rho_\text{A}\rho_\text{B})=1-2P_\text{odd}\,,
\end{equation}
where $P_\text{odd}$ is the probability that $\boldsymbol{q}^j\in\{0,1\}^N$ has odd parity, where $\boldsymbol{q}_n^j=\boldsymbol{r}_{2n-1}^j \cdot \boldsymbol{r}_{2n}^j$ is a bit-wise AND of the outcomes of each subsystem~\cite{garcia2013swap}. 
The SWAP test on two copies of the same state $\rho\otimes\rho$ can give us the purity $\text{tr}(\rho^2)$ as well as the entanglement $2$-Rényi entropy $\text{tr}(\rho^2_k)$ over a subsystem $\rho^2_k$ (see Appendix~\ref{sec:entangle}).

Now, we perform the Bell measurement on two copies of a pure state $\ket{\psi}\otimes\ket{\psi}$. The outcome $\boldsymbol{r}$ appears with a probability~\cite{montanaro2017learning}
\begin{equation}\label{eq:probBell}
P(\boldsymbol{r})=\bra{\psi}\bra{\psi}O_{\boldsymbol{r}}\ket{\psi}\ket{\psi}=2^{-N}\vert\bra{\psi}\sigma_{\boldsymbol{r}}\ket{\psi^*}\vert^2\,,
\end{equation} 
where $O_{\boldsymbol{r}}=\ket{\sigma_{\boldsymbol{r}}}\bra{\sigma_{\boldsymbol{r}}}$ is the projector onto a product of Bell states and 
$\ket{\psi^*}$ denotes the complex conjugate of $\ket{\psi}$. For any state, we have $0\le P(\boldsymbol{r})\le2^{-N}$ and there are between $2^N$ and $4^N$ outcomes $\boldsymbol{r}$ with $P(\boldsymbol{r})>0$.
For any set of bitstrings $\{\boldsymbol{r}^j\}_{j=1}^{N_\text{Q}}$ sampled in the Bell basis from a pure stabilizer state $\ket{\psi_\text{STAB}}\in G$, the Pauli strings of its binary additions must commute, i.e. $[\sigma_{\boldsymbol{r}^k\oplus\boldsymbol{r}^l},\sigma_{\boldsymbol{r}^n\oplus\boldsymbol{r}^m}]=0$ $\forall k,l,n,m$ (see Appendix~\ref{sec:Stabcommute} or \cite{montanaro2017learning}). Conversely, finding at least one non-commuting Pauli string implies that the measured quantum state is not a pure stabilizer state as the commuting subgroup $G$ contains at most $2^N$ elements. This motivates that the probability of observing non-commuting Pauli strings is a measure of distance to the set of pure stabilizer states. 

\begin{table}[htbp]\centering
\begin{tabular}{ |l|c| } 
\hline
Name&Symbol  \\
\hline
Bell magic & $\mathcal{B}$\\
Additive Bell magic &$\mathcal{B}_\text{a}$\\
Probability of outcome $\boldsymbol{r}\in\{0,1\}^{2N}$&$P(\boldsymbol{r})$\\
Estimation error  &$\Delta \mathcal{B}$\\
Pauli string&$\sigma_{\boldsymbol{n}}$\\
$\frac{1}{\sqrt{2}}(\ket{0}+e^{-i\frac{\pi}{4}}\ket{1})$&$\ket{T}$\\
$\cos(\frac{\theta}{2})\ket{0}+e^{-i\frac{\pi}{4}}\sin(\frac{\theta}{2})\ket{1}$, $\theta=\arccos(\frac{1}{\sqrt{3}})$&$\ket{R}$\\
$\cos(\frac{\phi}{2})\ket{0}+\sin(\frac{\phi}{2})\ket{1}$&$\ket{A_\phi}$\\
Number of qubits &$N$\\
Number of measurements &$N_\text{Q}$\\
Resampling steps &$N_\text{R}$\\
Number of magic states &$N_\text{A}$\\
Number of $T$-gates &$N_\text{T}$\\
Depolarizing error&$p$\\
Classification error probability&$P_\text{E}$\\
\hline
\end{tabular}
\caption{Definitions of symbols.}
\label{tab:definitions}
\end{table}

\sectionMain{Bell Magic}\label{sec:bellmagic}

We now define Bell magic $\mathcal{B}$ as 
\begin{equation}\label{eq:BellMagic}
\mathcal{B}=\sum_{\substack{\boldsymbol{r},\boldsymbol{r'},\boldsymbol{q},\boldsymbol{q'}\\\in\{0,1\}^{2N}}}P(\boldsymbol{r})P(\boldsymbol{r'})P(\boldsymbol{q})P(\boldsymbol{q'})\left\lVert[\sigma_{\boldsymbol{r}\oplus\boldsymbol{r'}},\sigma_{\boldsymbol{q}\oplus\boldsymbol{q'}}]\right\rVert_{\infty}
\end{equation}
where the infinity norm is zero $\lVert[\sigma_{\boldsymbol{r}},\sigma_{\boldsymbol{q}}]\rVert_{\infty}=0$ when the two Pauli strings commute $[\sigma_{\boldsymbol{r}}$, $\sigma_{\boldsymbol{q}}]=0$, and $\lVert[\sigma_{\boldsymbol{r}},\sigma_{\boldsymbol{q}}]\rVert_{\infty}=2$ otherwise. 
As a measure of magic~\cite{veitch2014resource}, $\mathcal{B}$ is faithful with $\mathcal{B}(\ket{\psi_\text{STAB}})=0$ only for pure stabilizer states $\ket{\psi_\text{STAB}}$, and $\mathcal{B}>0$ otherwise. 
As shown in Appendix~\ref{sec:invariance}, $\mathcal{B}$ is also invariant under Clifford circuits $U_C$ that map stabilizers to stabilizers, i.e. $\mathcal{B}(U_C\ket{\psi})=\mathcal{B}(\ket{\psi})$.
\revA{Further, Bell magic is constant under composition with any stabilizer state $\ket{\psi_\text{STAB}}$, i.e. $\mathcal{B}(\ket{\psi}\otimes\ket{\psi_\text{STAB}})=\mathcal{B}(\ket{\psi})$ (see Appendix~\ref{sec:composition}).
We numerically tested an extensive number of states and found that in all cases Bell magic is on average non-increasing under measurements in the computational basis over a set of qubits, i.e. $\mathcal{B}(\ket{\psi})\ge \sum_n q_n\mathcal{B}(\mathcal{M}_n(\ket{\psi}))$, where $\Pi_n=\ket{n}\bra{n}\otimes I_{N-1}$ is a projector on computational basis state $\ket{n}$, $I_{N-1}$ the identity operator for $N-1$ qubits, $q_n=\bra{\psi}\Pi_n\ket{\psi}$ the measurement probability and $\mathcal{M}_n(\ket{\psi})=q_n^{-1/2}\Pi_n \ket{\psi}$ the projected state.
}

We further define the additive Bell magic \begin{equation}\label{eq:magic_add}
\mathcal{B}_\text{a}=-\log_2(1-\mathcal{B})\,.
\end{equation}
$\mathcal{B}_\text{a}$ shares all properties with $\mathcal{B}$ and further is additive with $\mathcal{B}_\text{a}(\ket{\psi}\otimes\ket{\phi})=\mathcal{B}_\text{a}(\ket{\psi})+\mathcal{B}_\text{a}(\ket{\phi})$ which is proven in Appendix~\ref{sec:proofadditive}.
$\mathcal{B}_\text{a}$ has the operational meaning as the number of initial magic states $\ket{T}=T\ket{+}=\frac{1}{\sqrt{2}}(\ket{0}+e^{-i\frac{\pi}{4}}\ket{1})$ within any Clifford circuit $U_\text{C}$. For a state $\ket{T}^{\otimes k}\otimes\ket{0}^{\otimes (N-k)}$ consisting of a tensor product of $k$ magic states and else the stabilizer state $\ket{0}$,  additive Bell magic is given by
\begin{equation}
\mathcal{B}_\text{a}(U_C\ket{T}^{\otimes k}\otimes\ket{0}^{\otimes (N-k)})=k\,.
\end{equation}
The additive Bell magic of an $N$ qubit product state $\ket{\psi_\text{sp}(\boldsymbol{\theta},\boldsymbol{\varphi})}=\bigotimes_{n=1}^N(\cos(\frac{\theta_n}{2})\ket{0}+e^{-i\varphi_n}\sin(\frac{\theta_n}{2})\ket{1})$ is given by
\begin{align*}\label{eq:productBell}
\mathcal{B}_\text{a}(\boldsymbol{\theta},\boldsymbol{\varphi})=&-\sum_{n=1}^N\log_2[1-\frac{1}{32}\sin^2(\theta_n)(35+28\cos(2\theta_n)+\\
&\cos(4\theta_n)-8\cos(4\varphi_n)\sin^4(\theta_n))]\,,\numberthis
\end{align*}
which becomes maximal with $B_\text{a}^\text{R}=N\log_2(\frac{27}{11})\approx1.3N$ for the magic state $\ket{R}^{\otimes N}=(\cos(\frac{\theta}{2})\ket{0}+e^{-i\frac{\pi}{4}}\sin(\frac{\theta}{2})\ket{1})^{\otimes N}$ with $\theta=\arccos(\frac{1}{\sqrt{3}})$. For $N=1$ the state of maximal magic is $\ket{R}$ and for $N=3$  the Hoggar state, coinciding with the states of maximal robustness of magic~\cite{howard2017application}. We report the explicit forms of the pure states of maximal magic up to $N=4$ in Appendix~\ref{sec:maxmagic}. 
\revA{For pure states, we find that Bell magic is upper bounded by
\begin{equation}
    \mathcal{B}^\text{pure}\le4^{N}\frac{(1+2^{-N}-2\cdot4^{-N})^2}{(4^{N}-1)(1+2^{-N})^2}\,.
\end{equation}
This bound is not tight, but we find that it is saturated for $N=1$ and $N=3$. } 

\revA{Note as Bell magic is a measure of distance to the set of pure stabilizer states, it is in general non-zero for probabilistic mixtures of stabilizer states and thus not a proper measure of magic for generic mixed states. For example, the maximally mixed state $\rho_\text{m}=I_{N}2^{-N}$ with the $N$-qubit identity $I_{N}$ (which can be written as a probabilistic mixture of pure stabilizer states), has the maximal Bell magic with $\mathcal{B}(\rho_\text{m})=1-4^{-N}$ and $\mathcal{B}_\text{a}(\rho_\text{m})=2N$ (see Appendix~\ref{sec:mixed})

However, we can define an extension of Bell magic which is indeed faithful for a class of mixed stabilizer states.
We regard $N$-qubit mixed stabilizer states of the form $\rho_\text{STAB}=U_\text{C}\ket{\psi_\text{STAB}}\bra{\psi_\text{STAB}}\otimes I_K 2^{-K}U_\text{C}^\dagger$, where $\ket{\psi_\text{STAB}}$ is a $N-K$ qubit pure stabilizer state and $U_\text{C}$ an arbitrary $N$-qubit Clifford circuit. These states can be written as $\rho_\text{STAB}=2^{-N}(I+\sum_{\sigma\in G_0}\alpha_\sigma \sigma)$ where $\alpha_\sigma=\pm1$ and $G_0\subseteq G/\{I_N\}$ is a subset of a commuting subgroup of Pauli strings (excluding identity $I_N$) with $\vert G_0\vert\le 2^{N}-1$. We define the mixed Bell magic as
\begin{equation}\label{eq:mixedmagic}
\mathcal{B}_\text{m}(\rho)=1-\frac{1-\mathcal{B}(\rho)}{\text{tr}(\rho^2)^2}\,.
\end{equation}
$\mathcal{B}_\text{m}$ shares all properties of $\mathcal{B}$ and additionally we have $\mathcal{B}_\text{m}(\rho_\text{STAB})=0$ (see Appendix~\ref{sec:mixedmagic}). We numerically checked various states and found that in all cases $\mathcal{B}_\text{m}$ is non-increasing when partially tracing out qubits.
We also define the additive mixed Bell magic 
\begin{equation}\label{eq:mixedmagic_add}
\mathcal{B}_\text{a,m}(\rho)=-\log_2(1-\mathcal{B}(\rho))+2\log_2(\text{tr}(\rho^2))\,.
\end{equation}
Note that we can also use error mitigation to extract Bell magic from noisy states, which we show in Sec.\ref{sec:errormitigation}.}

\sectionMain{Theory of measuring magic}\label{sec:measuremagic}

\begin{algorithm}[h]
 \SetAlgoLined
 \LinesNumbered
  \SetKwInOut{Input}{Input}
  \SetKwInOut{Output}{Output}
   \Input{Bitstrings $r,q\in\{0,1\}^{2N}$
   }
    \Output{$C$}
    $\sigma_{\boldsymbol{r}}=\bigotimes_{n=1}^N\boldsymbol{\sigma}_{\boldsymbol{r}_{2n-1}\boldsymbol{r}_{2n}}$
    
    $\sigma_{\boldsymbol{q}}=\bigotimes_{n=1}^N\boldsymbol{\sigma}_{\boldsymbol{q}_{2n-1}\boldsymbol{q}_{2n}}$
    
    \eIf{$[\sigma_{\boldsymbol{r}},\sigma_{\boldsymbol{q}}]=0$}
    {$ C\leftarrow 0 $}
    {$ C\leftarrow 2 $}

 \caption{Check-Commute}
 \label{alg:checkcommute}
\end{algorithm}

\begin{algorithm}[h]
 \SetAlgoLined
 \LinesNumbered
  \SetKwInOut{Input}{Input}
  \SetKwInOut{Output}{Output}
   \Input{$j=1,\dots,N_\text{Q}$ bitstrings $r^j\in\{0,1\}^{2N}$ sampled from Bell measurement\\
   Resampling steps $N_\text{R}$}
    \Output{Bell magic $\mathcal{B}$\\
    Additive Bell magic $\mathcal{B}_\text{a}$
    }

$\mathcal{B}\leftarrow 0$

 \SetKwRepeat{Do}{do}{while}
    \For{$k=1,\dots,N_\text{R}$}{
    Choose randomly without replacement $n_1,n_2,n_3,n_4\in\{1,\dots,N_\text{Q}\}$
    $\mathcal{B}\leftarrow \mathcal{B}+\text{Check-Commute}(r^{n_1}\oplus r^{n_2},r^{n_3}\oplus r^{n_4})$
    }
    $\mathcal{B}\leftarrow \mathcal{B}/N_\text{R}$
    
    $\mathcal{B}_\text{a}\leftarrow -\log_2(1-\mathcal{B})$
 \caption{Bell magic}
 \label{alg:bellmagic}
\end{algorithm}

Bell magic $\mathcal{B}$ can be efficiently estimated from measurements on quantum states.
We give an unbiased estimator for $\mathcal{B}$ in Algorithm~\ref{alg:checkcommute} and Algorithm~\ref{alg:bellmagic}. \revA{We prepare two copies of the state, perform a Bell measurements between them, and record the outcome. We repeat this step $N_\text{Q}$ times, requiring in total $2N_\text{Q}$ copies of the state.} Then, we post-process the outcomes. We randomly draw four bitstrings from the outcomes  without replacement and check whether their addition commutes. This step is repeated for $N_\text{R}$ times, where $N_\text{R}$ are the resampling steps and we always draw the bitstrings from all $N_\text{Q}$ outcomes. $\mathcal{B}$ is then estimated as the probability of getting a non-commuting result.

\revA{As we will show in the next paragraph, the number of measurements $N_\text{Q}$ needed to estimate $\mathcal{B}$ with fixed accuracy is independent of qubit number $N$, i.e. $N_\text{Q}\sim O(1)$. Further, the classical post-processing scales as $O(N)$.  On first glance this seems counter-intuitive, as the number of possible outcomes scales exponentially with $N$. However, to estimate $\mathcal{B}$ we assign the bitstrings only two possible values via the Check-commute routine (corresponding to the norm of the commutator in \eqref{eq:BellMagic}). Bell magic is then estimated as the expectation value of the two values. Thus, the measurement process corresponds to a Bernoulli trial, with the same scaling of errors as estimating the expectation value of a coin flip or a Pauli operator.  }

Now, we give analytic bounds on the estimation error of $\mathcal{B}$ for the case $N_\text{R}= N_\text{Q}/4$. We slightly modify Algorithm~\ref{alg:bellmagic} such that the bitstrings are not sampled at random, but each of the $N_\text{Q}$ bitstrings is drawn exactly once. 
As we use each bitstring only once, the outcomes of the Check-Commute routine are statistically independent and we can write the estimated Bell magic as a Bernoulli trial $\hat{\mathcal{B}}=2p_\text{non-cm}=2M_\text{non-cm}/M_\text{total}$, where $p_\text{non-cm}$ is the probability that an outcome does not commute, $M_\text{non-cm}$ is the number of outcomes that do not commute and $M_\text{total}=N_\text{Q}/4$ is the total number of repetitions. 
The standard deviation of the Bell magic for such Bernoulli experiments is given by 
\begin{equation}\label{eq:std_Bernoulli}
\text{std}(\mathcal{B})=2\sqrt{\frac{p_\text{non-cm}(1-p_\text{non-cm})}{M_\text{total}}}=\sqrt{\frac{8\mathcal{B}}{N_\text{Q}}(1-\frac{\mathcal{B}}{2})}\,.
\end{equation}
The number of measurement samples $N_\text{Q}$ needed to achieve an estimation error of at most $\Delta \mathcal{B}$ with a failure probability $P_\text{F}$ is bounded by Hoeffding's inequality
\begin{equation}
    P(\vert \hat{\mathcal{B}} -\mathcal{B}\vert\ge \Delta \mathcal{B})=P_\text{F}\le 2\exp(-\frac{2\Delta \mathcal{B}^2 M_\text{total}}{(a_\text{max}-a_\text{min})^2})\,,
\end{equation}
where $a_\text{max}=2$ and $a_\text{min}=0$ are the maximal and minimal possible values of each trial of Algorithm~\ref{alg:checkcommute}.
The upper bound for the needed samples is given by
\begin{equation}\label{eq:sample_B}
N_\text{Q}\ge\frac{8}{\Delta\mathcal{B}^2}\log(\frac{2}{P_\text{F}})\,.
\end{equation}
In particular, the estimation error scales as $\Delta\mathcal{B}\propto N_\text{Q}^{-\frac{1}{2}}$ and is independent of qubit number $N$.
Above equations are derived for the choice $N_\text{R}= N_\text{Q}/4$. By increasing the number of post-processing steps $N_\text{R}>N_\text{Q}/4$ the accuracy increases further. We numerically find that $N_\text{R}=10N_\text{Q}$ provides estimates of the Bell magic close to the maximal possible accuracy. Thus, we find that the classical post-processing has $O(N)$ complexity in time and memory. 

\revA{The outcomes of the Bell measurement also allow us to compute the purity $\text{tr}(\rho^2)$ via~\eqref{eq:SWAP} at the same time. The purity is estimated as the probability of measuring outcomes of odd parity, which again is a Bernoulli trial. The estimation error of the purity scales as $\propto N_\text{Q}^{-\frac{1}{2}}$, which allows us to also estimate mixed Bell magic (~\eqref{eq:mixedmagic}) efficiently.}

\sectionMain{Error mitigation}\label{sec:errormitigation}

Next, we use the purity to mitigate errors from the outcomes of noisy quantum computers~\cite{endo2021hybrid,vovrosh2021efficient,haug2021large}. Our goal is to determine the Bell magic of the pure state $\ket{\psi}$ by measuring the state $\rho_\text{dp}=(1-p)\ket{\psi}\bra{\psi}+p\rho_m$ subject to global depolarizing noise with a probability $p$~\cite{nielsen2002quantum}. Depolarizing noise has been shown to be a good approximation in experiments on noisy quantum computers~\cite{vovrosh2021simple,haug2021large} and coherent errors can be turned into depolarizing errors via randomized compiling~\cite{wallman2016noise,urbanek2021mitigating}. As will be seen in Sec.~\ref{sec:results}, our experimental results are well described with a depolarizing model.
We prepare two copies $\rho_\text{dp}\otimes\rho_\text{dp}$ and apply the Bell transformation, where we assume that the Bell measurement is noise-free. From the Bell measurements, we determine the purity $\text{tr}(\rho_\text{dp}^2)$ via~\eqref{eq:SWAP} as well as the Bell magic $\mathcal{B}^\text{dp}$ of the noise-affected state $\rho_\text{dp}$. 
The purity is related to the depolarization error via $\text{tr}(\rho_\text{dp}^2)=(1-p)^2+\frac{p(2-p)}{2^N}$. By inverting we get the depolarization error 
\begin{equation}\label{eq:depolpurity}
p=1-\frac{\sqrt{(2^N-1)(2^N\text{tr}(\rho_\text{dp}^2)-1)}}{2^N-1}\,.
\end{equation}
The mitigated Bell $\mathcal{B}^\text{mtg}$ of the noise-free state $\ket{\psi}$ is given by (see Appendix~\ref{sec:error_mtg_sup})
\begin{equation}\label{eq:BellMtg}
\mathcal{B}^\text{mtg}=\frac{1}{(1-p_\text{c})^2}(\mathcal{B}^\text{dp}-p_\text{c}^2\mathcal{B}(\rho_\text{m})-2p_\text{c}(1-p_\text{c})\mathcal{B}^\text{R})\,,
\end{equation}
where $p_\text{c}=1-(1-p)^4$,  $\mathcal{B}^\text{R}=1-(1-p_\text{c})^{-1}(\sum_{\boldsymbol{q}}P_\text{dp}(\boldsymbol{q})^2-4^{-N}p_\text{c})$. Here, $P_\text{dp}(\boldsymbol{q})$ is the probability of measuring bitstring $\boldsymbol{q}$ of the noisy state. 
In the limit of many qubits $N$, we approximate $\mathcal{B}^\text{R}\approx \mathcal{B}(\rho_\text{m})\approx1$ and get
\begin{equation}\label{eq:Bellmtgapprox}
\mathcal{B}^\text{mtg}\approx\frac{\mathcal{B}^\text{dp}-p_\text{c}(2-p_\text{c})}{(1-p_\text{c})^2}\,.
\end{equation}
We now give the scaling of the number of samples $N_\text{Q}$ needed to estimate the mitigated magic with an error $\Delta \mathcal{B}$. 
We define the error $\Delta\mathcal{B}^\text{dp}=\vert\hat{\mathcal{B}}^\text{dp}-\mathcal{B}^\text{dp}\vert$ of $\hat{\mathcal{B}}^\text{dp}$ estimated by measuring the noisy quantum computer. We insert \eqref{eq:Bellmtgapprox} and get the error of the mitigated magic
$\Delta\mathcal{B}^\text{dp}\approx(1-p)^{8}\Delta\mathcal{B}^\text{mtg}$,
where we used $(1-p_\text{c})^2=(1-p)^8$. The upper bound of $\Delta\mathcal{B}^\text{dp}$ is given by \eqref{eq:sample_B}, where we insert the mitigated error $\Delta\mathcal{B}^\text{mtg}$. Note that the upper bound can be slightly violated to our approximations and the error in the estimation of $p$. However, we argue that the scaling of the error remains the same which we confirm numerically.
Thus, the number of samples needed to estimate the mitigated magic within error $\Delta\mathcal{B}^\text{mtg}$ scales as \begin{equation}\label{eq:errormtg}
N_\text{Q}\propto \frac{1}{(1-p)^{16}{\Delta\mathcal{B}^\text{mtg}}^2}\,.
\end{equation}

\sectionMain{Demonstration of measuring Bell magic}\label{sec:results}

\begin{figure*}[htbp]
	\centering	
		\subfigimg[width=0.33\textwidth]{a}{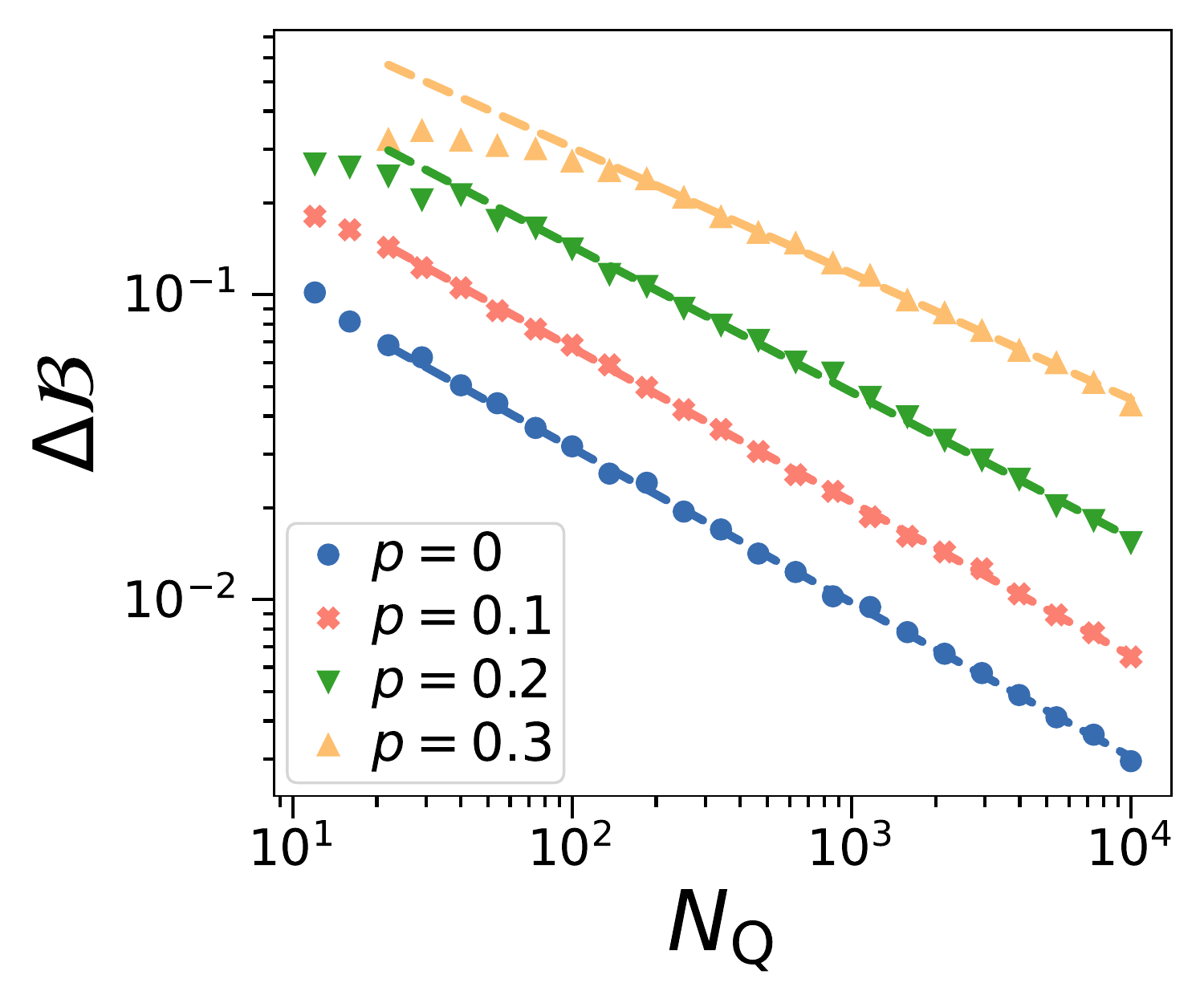}\hfill
	\subfigimg[width=0.33\textwidth]{b}{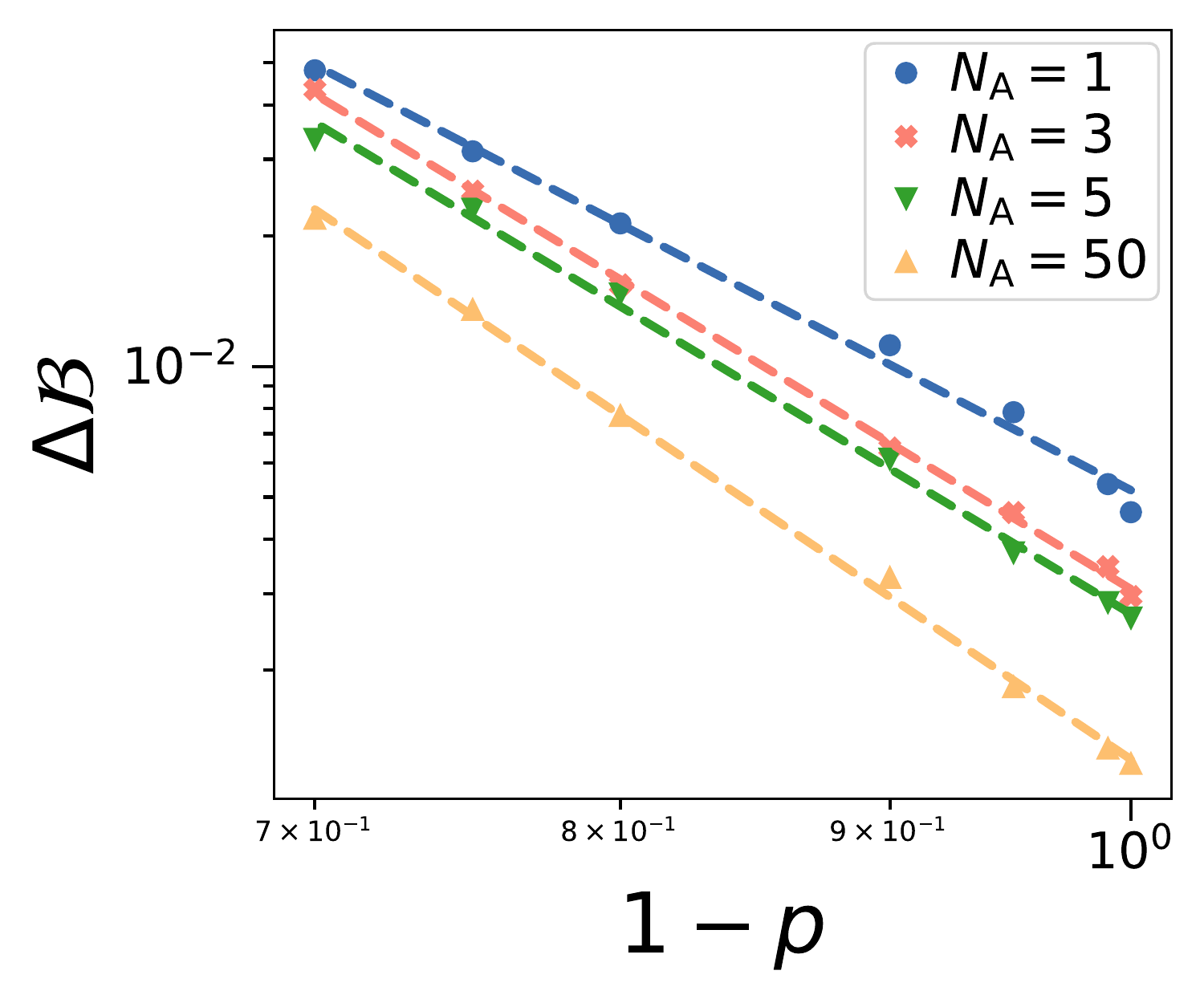}\hfill
		\subfigimg[width=0.33\textwidth]{c}{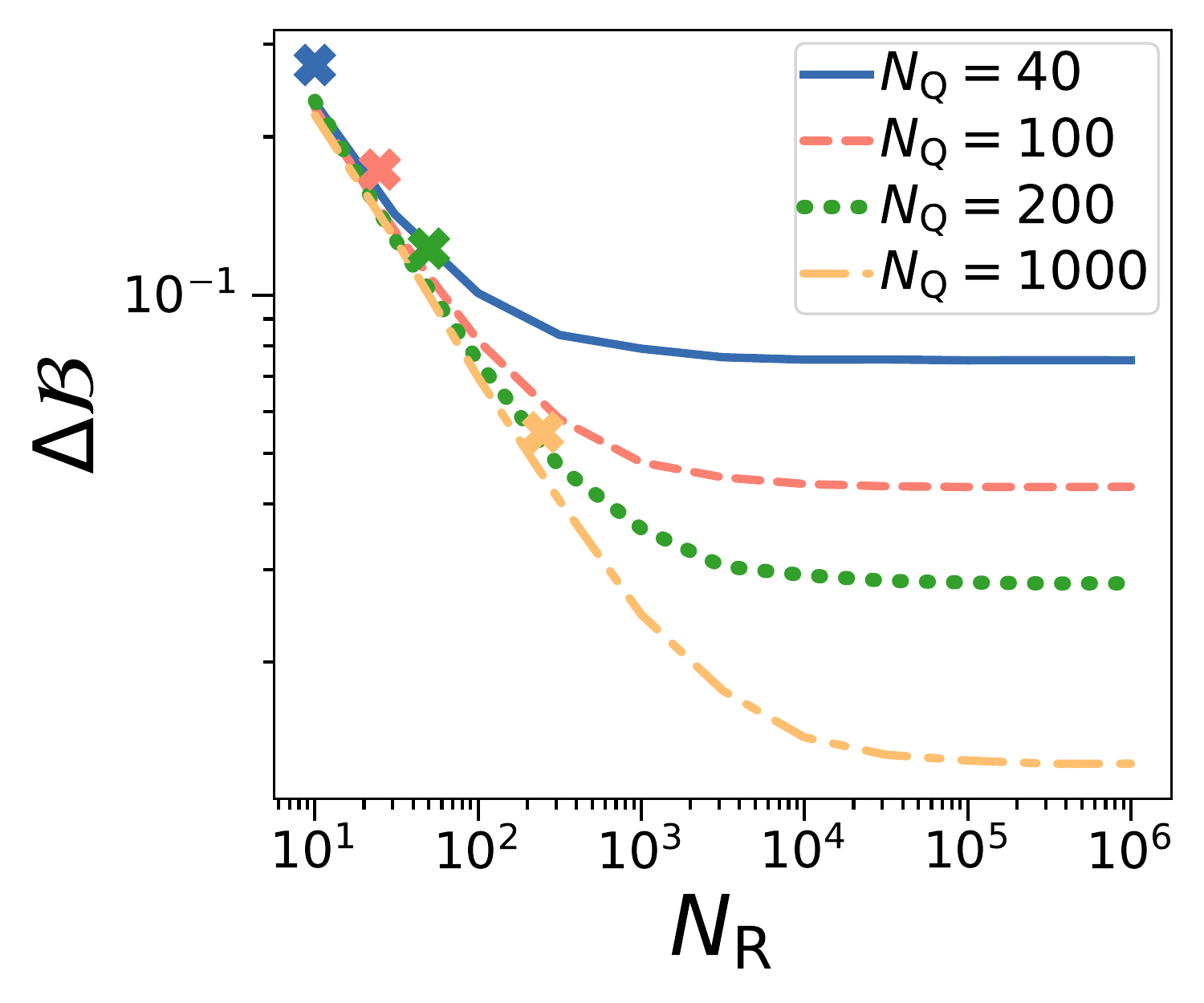}
	\caption{\idg{a} Simulation of estimation error of mitigated Bell magic $\Delta \mathcal{B}=\langle\vert\hat{\mathcal{B}}^\text{mtg}-\mathcal{B}^\text{exact}\vert\rangle$ as a function of the number of measurements $N_\text{Q}$ for varying depolarizing probability $p$. The dashed lines are linear fits with a slope of $b=\{-0.507, -0.505, -0.479, -0.412\}$ in descending order of the legend. We use a random Clifford circuit of $N=50$ qubits applied on an initial state of $N_\text{A}=3$ magic states with $\phi=\pi/4$ and $N_\text{R}=10N_\text{Q}$. The error is averaged over 1000 repetitions.
	\idg{b} $\Delta \mathcal{B}$ against depolarizing error $p$ for varying number of magic states $N_\text{A}$ and random $U_\text{C}$. Dashed lines are fits with slope $b=\{-6.32, -7.36, -7.307, -8.18\}$. We use $N=50$, $N_\text{Q}=10^4$ and $N_\text{R}=10N_\text{Q}$.
	\idg{c} $\Delta \mathcal{B}$ as a function of classical resampling steps $N_\text{R}$ after performing $N_\text{Q}$ measurements on a quantum computer. Crosses are the theoretical values of the standard deviation \eqref{eq:std_Bernoulli} for $N_\text{R}=N_\text{Q}/4$. We use a random Clifford circuit of $N=8$ qubits applied on an initial state of $N_\text{A}=1$ magic states.
	}
	\label{fig:measure}
\end{figure*}

Now, we demonstrate the measurement of Bell magic. First, we numerically investigate in Fig.\ref{fig:measure} the dependence of the estimation error on various parameters. We measure states $U_\text{C}\ket{A_\phi}^{\otimes N_\text{A}}\otimes\ket{0}^{N-N_\text{A}}$ with $N=50$ qubits consisting of randomly chosen Clifford circuits $U_\text{C}$ applied to a product of $N_\text{A}$ the state $\ket{A_\phi}=\cos(\frac{\phi}{2})\ket{0}+\sin(\frac{\phi}{2})\ket{1}$ with angle $\phi$, which we realize as hardware efficient quantum circuits~\cite{haug2021capacity} (see Appendix~\ref{sec:hardwareefficient} for details). \revA{We have $\mathcal{B}_\text{a}(\ket{A_{\phi=0}})=0$ and $\mathcal{B}_\text{a}(\ket{A_{\phi=\pi/4}})=\mathcal{B}_\text{a}(\ket{T})=1$.}
We numerically simulate states of $N=50$ qubits with tensor network methods~\cite{itensor,pastaq}.
In Fig.\ref{fig:measure}a we plot the estimation error $\Delta \mathcal{B}=\langle\vert\hat{\mathcal{B}}^\text{mtg}-\mathcal{B}^\text{exact}\vert\rangle$ between the mitigated magic $\hat{\mathcal{B}}^\text{mtg}$ estimated from $N_\text{Q}$ measurements and the exact value $\mathcal{B}^\text{exact}$. 
We show the error for varying number of Bell measurements $N_\text{Q}$ and depolarizing error $p$.
We find that the fit matches the prediction $\Delta\mathcal{B}\propto N_\text{Q}^{-1/2}$ (\eqref{eq:errormtg}). We find a slightly different slope for large $p$ which could be the result of not including estimation errors in the purity into our theory.
In Fig.\ref{fig:measure}b we plot $\Delta \mathcal{B}$  against depolarizing error $p$ for varying number of magic states $N_\text{A}$ within a Clifford circuit. The fit to the data is close to the relation $\Delta \mathcal{B}\propto (1-p)^{-8}$ of \eqref{eq:errormtg}.
In Fig.\ref{fig:measure}c we plot $\Delta \mathcal{B}$ as a function of resampling steps $N_\text{R}$ for various number of measurements $N_\text{Q}$. We find that the theoretical model \eqref{eq:std_Bernoulli} matches our simulation accurately.
Increasing $N_\text{R}>N_\text{Q}/4$ improves the accuracy until it converges for large $N_\text{R}$ to a constant value. Our numerics suggest that $N_\text{R}=10N_\text{Q}$ is a good empirical choice that gives nearly the lowest possible error.

\begin{figure}[htbp]
	\centering	
	\subfigimg[width=0.45\textwidth]{}{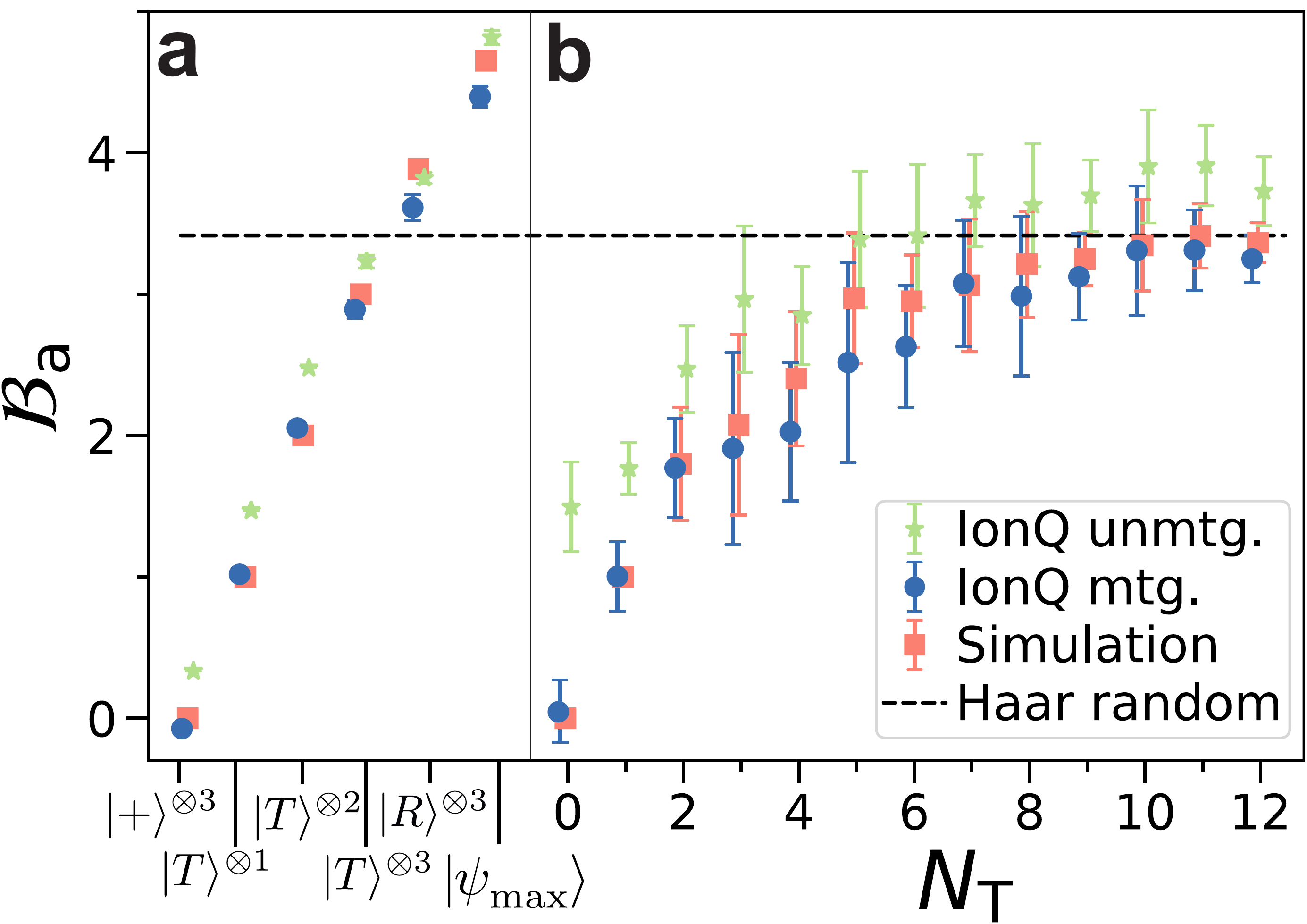}
	\caption{Experiment to measure the additive Bell magic $\mathcal{B}_\text{a}$ on the IonQ quantum computer for various types of states. \idg{a} In the left part of the graph, we show product states of stabilizer states $\ket{+}^{\otimes N}$, magic states $\ket{T}^{\otimes N}$, $\ket{R}^{\otimes N}$ as well as the state of maximal magic $\ket{\psi_\text{max}}$ for $N=3$. \idg{b} In the right part of the graph, we show magic as a function of $N_\text{T}$ $T$-gates inserted at random positions in a Clifford circuit.  
	We show the unmitigated and mitigated magic from IonQ quantum computer and an exact simulation of the quantum states. The mean value and the standard deviation of $\mathcal{B}_\text{a}$ is taken over 6 random instances of the states for $N=3$ qubits. The dashed line is the additive Bell magic averaged over Haar random states. The experiment is performed with $N_\text{Q}=10^3$ measurement samples and no further error or readout error mitigation. With the purity measured from the experiment, \eqref{eq:depolpurity} gives on average a depolarization error of $p\approx0.1$ for the IonQ quantum computer. 
	}
	\label{fig:ionqmeas}
\end{figure}

Next, we experimentally measure in Fig.\ref{fig:ionqmeas} the magic of various states on the 11-qubit IonQ quantum computer~\cite{wright2019benchmarking}. We prepare two instances of the desired quantum state on the quantum computer, then apply the Bell transformation and measure each qubit. 
We investigate the additive Bell magic of various types of states. \revA{In particular, we measure different product states and the state of maximal Bell magic in Fig.\ref{fig:ionqmeas}a, as well as stabilizer states with a variable amount of injected $T$-gates in Fig.\ref{fig:ionqmeas}b. We note that while product states are not entangled, the other states are substantially entangled, which we confirm experimentally with the Meyer-Wallach measure~\cite{meyer2002global} in Appendix~\ref{sec:entangle}.}
The error mitigation with \eqref{eq:BellMtg} substantially improves the results, with the mitigates $\mathcal{B}_\text{a}$ matching the exact simulations quite well. In Fig.\ref{fig:ionqmeas}a we observe for $\ket{T}^{\otimes N}$ a linear increase in $\mathcal{B}_\text{a}$ with $N$, highlighting its additive property.
In Fig.\ref{fig:ionqmeas}b, we study the transition of classically simulable stabilizer states to intractable quantum states~\cite{leone2021quantum}.
We prepare a state of the form $U_\text{C}\prod_{n=1}^{N_\text{T}}U_\text{T}U_\text{C}^n\ket{0}$, where $U_\text{C}^n$ is a randomly chosen Clifford circuit and $U_\text{T}=I_2\otimes\dots\otimes T\otimes\dots\otimes I_2$ is a  $T$-gate placed at a random qubit. See Appendix~\ref{sec:hardwareefficient} for the practical implementation of the circuit on quantum computers. 
For $N_\text{T}=0$, we prepare a stabilizer state with $\mathcal{B}_\text{a}=0$. Non-stabilizer states are prepared for $N_\text{T}>0$. We find that $\mathcal{B}_\text{a}$ grows nearly monotonous with $N_\text{T}$ until it converges. 
We find that the converged $\mathcal{B}_\text{a}$ matches closely the Bell magic averaged over Haar random states.
Agreeing with our observations, recent results showed that with increasing $N_\text{T}$ circuits will closely approximate unitaries sampled from the distribution of Haar random states~\cite{brandao2016local,haferkamp2020quantum,haferkamp2022quantum}.

\sectionMain{State discrimination}\label{sec:discrimination}
The efficient measurement of magic opens up new applications in state discrimination. When performing quantum computation or communicating over quantum networks, an important task is to verify whether a given unknown state possesses the desired properties such as a sufficient amount of magic (see Fig.\ref{fig:sketch}c).
Now, our goal is to determine the correct class of an unknown state $\ket{\psi}$ sampled from one of two classes $\alpha$, $\beta$ with different Bell magic $\mathcal{B}(\ket{\psi}\in\alpha)=\mathcal{B}^\alpha$, $\mathcal{B}(\ket{\psi}\in\beta)=\mathcal{B}^\beta$ with $\mathcal{B}^\alpha>\mathcal{B}^\beta$.
To this end, we perform $N_\text{Q}$ repetitions of the Bell measurements and estimate $\hat{\mathcal{B}}$. We choose an appropriate threshold $\mathcal{B}^*$. For $\hat{\mathcal{B}}>\mathcal{B}^*$, we decide that the given state belongs to $\alpha$, while for 
$\hat{\mathcal{B}}\le\mathcal{B}^*$ we say the state belongs to $\beta$. We define $P_\text{E}$ as the probability of wrongly classifying a state in the state discrimination protocol.

We motivate now the scaling of the number of measurements $N_\text{Q}$ needed for the classification task with a misclassification probability $P_\text{E}$. If the estimation error $\Delta \mathcal{B}$ of $\mathcal{B}(\ket{\psi})$ is larger than $\mathcal{B}^\alpha-\mathcal{B}^\beta$, the estimation error is too large to reliably distinguish the two classes.
Thus, the estimation error must be smaller than the difference in magic of the two states $\Delta \mathcal{B}<\mathcal{B}^\alpha-\mathcal{B}^\beta$ to reliably distinguish the states.
\eqref{eq:errormtg} tells us how many measurements are needed to estimate magic with additive error $\Delta \mathcal{B}$. We argue that the classification task follows in general the same scaling as \eqref{eq:errormtg} in the number of samples $N_\text{Q}$
\begin{equation}\label{eq:errorClass}
N_\text{Q}\approxprop\frac{1}{(1-p)^{16}(\mathcal{B}^\alpha-\mathcal{B}^\beta)^2}\,.
\end{equation}
An important special case is discriminating magical states $\mathcal{B}^\alpha>0$ from stabilizer states with $\mathcal{B}^\beta=0$. For this case we can derive the precise number of measurements needed for a threshold $\mathcal{B}^*=0$ and $p=0$.

First, we study a state with low magic 
\begin{equation}\label{eq:low_magic_state}
\ket{\psi_C(\phi)}=U_C\ket{A_\phi}\otimes\ket{0}^{N-1}
\end{equation}
consisting of an arbitrary Clifford circuit $U_C$ and an initial state $\ket{A_\phi}\otimes\ket{0}^{N-1}$ with $N_\text{A}=1$ non-stabilizer qubit $\ket{A_\phi}=\cos(\frac{\phi}{2})\ket{0}+\sin(\frac{\phi}{2})\ket{1}$. Here, $\phi$ controls the amount of magic introduced into the circuit as seen in \eqref{eq:productBell}. In particular for $\phi=n\pi/2$, $n$ being an integer, no magic is introduced, whereas for $\phi=\pi/4$ we have $\mathcal{B}_\text{a}=1$. 
For small $\phi$, Bell magic can be approximated as $\mathcal{B}(\vert\phi\vert\ll1)\approx2\phi^2$. By tuning $\phi$, we can create states containing arbitrarily low amounts of magic. 
We define the error probability $P_\text{E}(\phi)$ as the probability of wrongly classifying $\ket{\psi_C(\phi)}$ as a stabilizer state. 
With the assumption of large $N_\text{R}$ we find as shown in Appendix~\ref{sec:singleqdiscr}
\begin{align*}
P_\text{E}(\phi)=& 4^{-N_\text{Q}}[(3-\cos(2\phi))^{N_\text{Q}}+(3+\cos(2\phi))^{N_\text{Q}}]-\\
&2^{-N_\text{Q}}[\sin(\phi)^{2N_\text{Q}}+\cos(2\phi)^{2N_\text{Q}}]\numberthis\label{eq:errorSingleT}
\end{align*}
For $\phi=\frac{\pi}{4}$, we can approximate the error probability as
\begin{equation}
P_\text{E}(\frac{\pi}{4})\approx2\left(\frac{3}{4}\right)^{N_\text{Q}}\,.
\end{equation}
We achieve an error $P_\text{E}(\frac{\pi}{4})<0.01$ when $N_\text{Q}>18$.
For near-stabilizer states with small $\phi\ll1$ we find
\begin{equation}
N_\text{Q}\approx-\frac{2\log(P_\text{E}(\phi\ll1))}{\phi^2}\,,
\end{equation}
showing an inverse quadratic scaling law with $\phi$.
For example, a modest budget of $N_\text{Q}=375$ samples is needed to classify $\phi=\frac{\pi}{20}$ with an error of $P_\text{E}^\text{min}<0.01$. Surprisingly, for small $\phi$ we find a scaling $N_\text{Q}\propto \mathcal{B}^{-1}$. This scaling is better than \eqref{eq:errorClass} which was derived with the additional assumption of a relatively small number of resampling steps $N_\text{R}=N_\text{Q}/4$.

As second case, we study the limit of highly magical states with $N_\text{A}\gg1$.
For these types of states we can assume that the $N_\text{Q}$ Bell measurements yield random bitstrings. As we add two bitstrings together in Algorithm~\ref{alg:bellmagic}, we have in total $N_\text{Q}-1$ independent bitstrings and their associated Pauli strings.
The probability that two random Pauli strings commute is $\approx\frac{1}{2}$.
The misclassification probability is given by the probability that $N_\text{Q}-1$ Pauli strings pairwise commute
\begin{equation}\label{eq:errorrandom}
P_\text{E}(N_\text{A}\gg1)\approx 2^{-(N_\text{Q}-1)(N_\text{Q}-2)/2}\,.
\end{equation}
In particular, we achieve an error probability $P_\text{E}(N_\text{A}\gg1)<0.01$ when $N_\text{Q}>5$.

\begin{figure}[htbp]
	\centering	
	\subfigimg[width=0.33\textwidth]{}{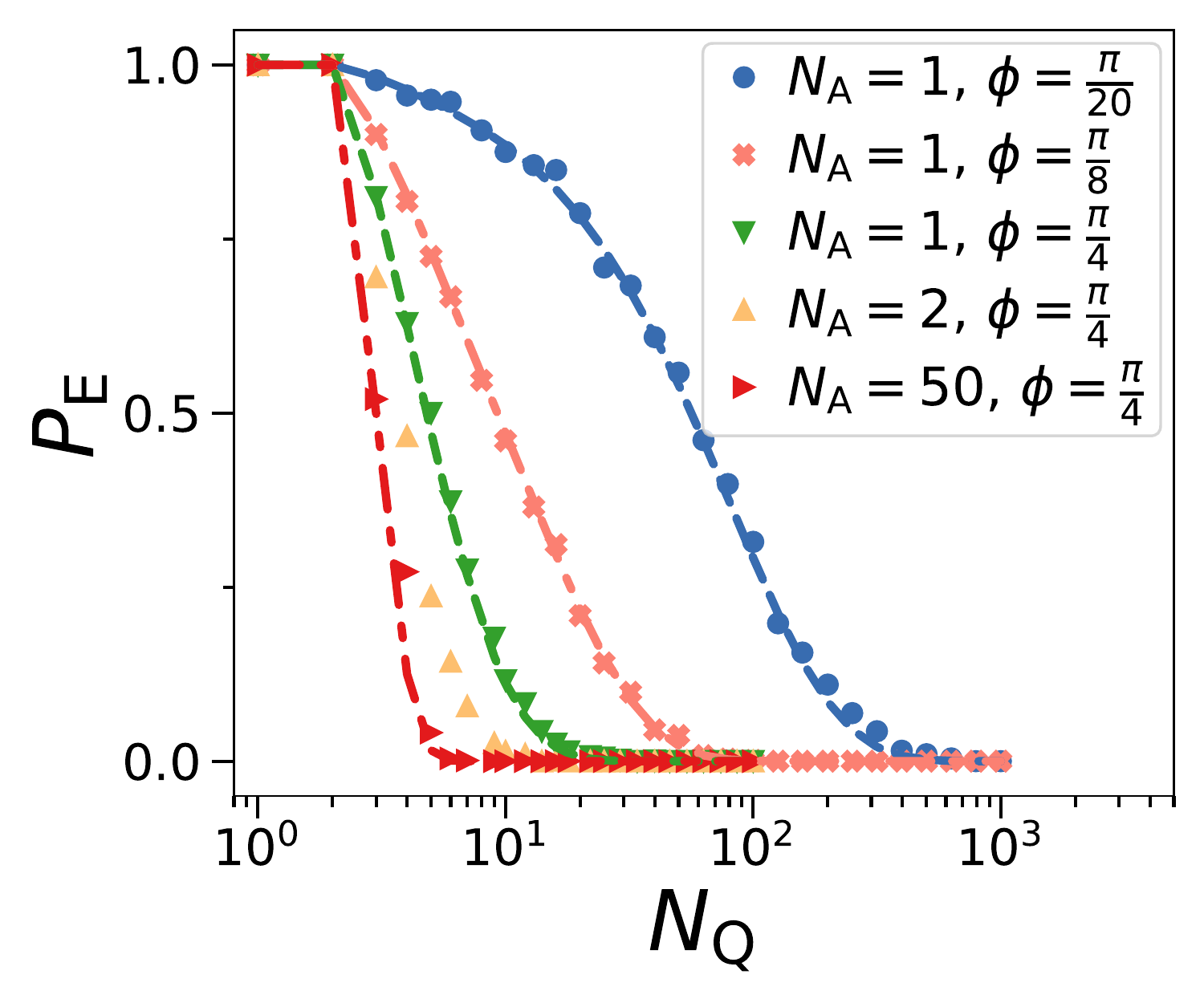}
	\caption{Simulation of error probability $P_\text{E}$ for wrongly classifying a given magical state as a stabilizer state using $N_\text{Q}$ measurements and threshold $\mathcal{B}^*=0$. The measured states are randomly chosen Clifford circuits of $N=50$ qubits applied to a product state of $N_\text{A}$ magic states where angle $\phi$ controls the amount of magic introduced, with $\phi=0$ zero magic and $\phi=\frac{\pi}{4}$ maximal magic. The dashed lines are \eqref{eq:errorSingleT} for the $N_\text{A}=1$ cases with $\phi=\frac{\pi}{20}$ (blue), $\phi=\frac{\pi}{8}$ (orange) and $\phi=\frac{\pi}{4}$ (green). The red dashed line \eqref{eq:errorrandom} is the analytic formula for highly magical states.
	}
	\label{fig:discriminate}
\end{figure}

Now, we numerically demonstrate our magic state discrimination protocol. In Fig.\ref{fig:discriminate} we simulate the states $\ket{\psi(\phi)}=U_C\ket{A_\phi}^{\otimes N_\text{A}}\otimes\ket{0}^{\otimes (N-N_\text{A})}$ where $U_C$ is a randomly chosen Clifford circuit and $N_\text{A}$ is the number of non-stabilizer input qubits. 
We measure the state $N_\text{Q}$ times and estimate the magic $\hat{\mathcal{B}}$ with Algorithm \ref{alg:bellmagic}. 
For $N_\text{Q}\le3$, we cannot draw the bitstrings without replacement in Algorithm~\ref{alg:bellmagic} and thus we draw with replacement instead. If we measure $\hat{\mathcal{B}}=\mathcal{B}^*=0$, we incorrectly say that the state is a stabilizer state, else for $\hat{\mathcal{B}}>0$ we classify the state as a non-stabilizer state.
We find that our numerical results for the error probability $P_\text{E}$ fit very well with our theoretical formulas.

\begin{figure}[htbp]
	\centering	
	\subfigimg[width=0.33\textwidth]{}{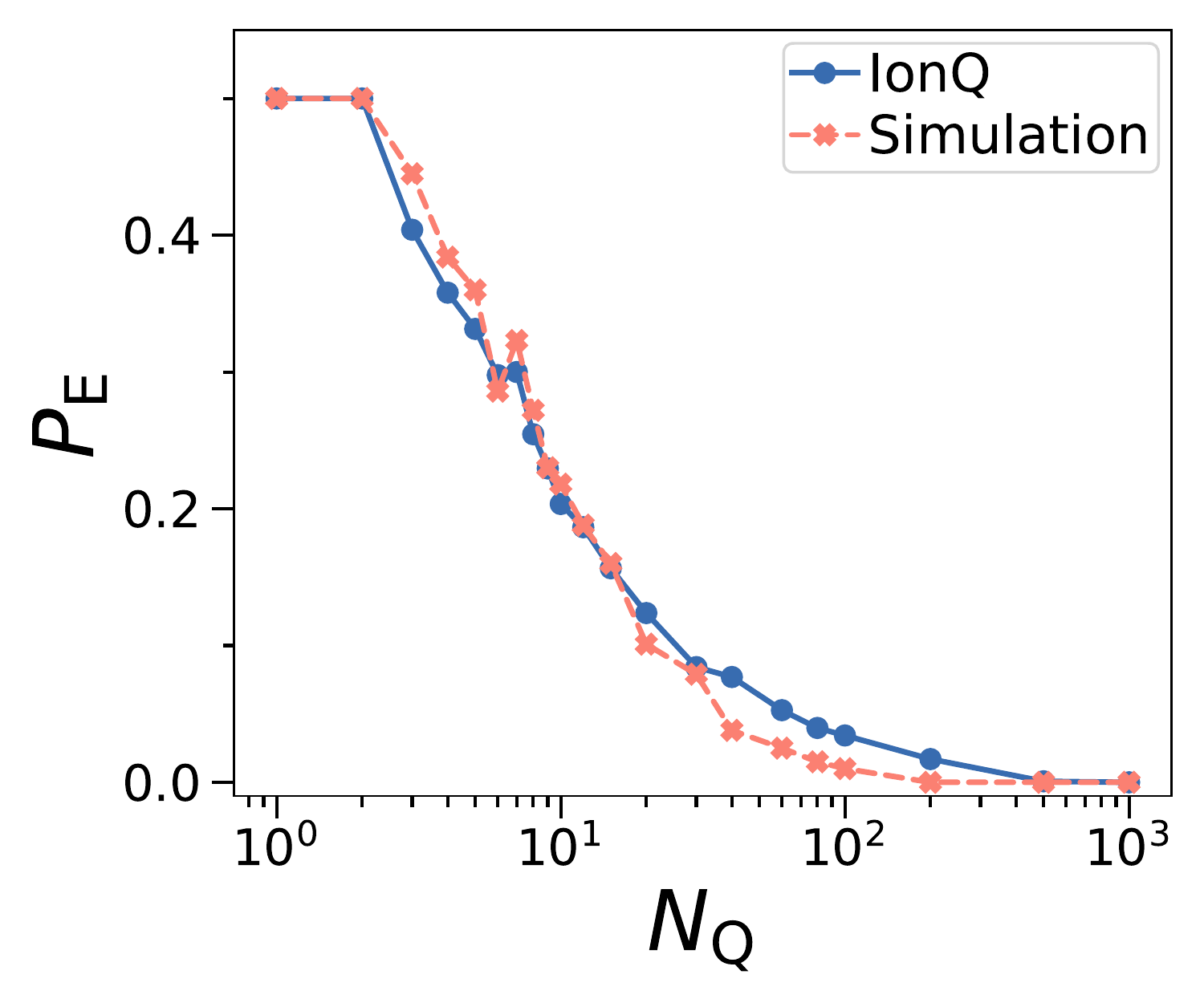}
	\caption{Experiment to distinguish stabilizer and highly magical states with the IonQ quantum computer using Bell magic. We show classification error $P_\text{E}$ as a function of the number of measurements $N_\text{Q}$. The supervised learning algorithm to determine the best threshold $\mathcal{B}^*$ is shown in Appendix~\ref{sec:suplearn}. We measure 20 randomly chosen instances of stabilizer and magical states respectively, which are generated with a hardware efficient circuit of $d=2$ depth and $N=3$ qubits. The shown curves are the classification error for the test dataset, which consists of 20\% of the data and is not seen during training. The error is averaged over 10 random distributions of test and training data.  Blue dots is experiment with IonQ quantum computer, while orange crosses are a noisy simulation using the average experimentally measured depolarization error $p\approx0.15$.
	}
	\label{fig:ionqlearn}
\end{figure}

Next, in Fig.\ref{fig:ionqlearn} we use the IonQ quantum computer to experimentally distinguish stabilizer and non-stabilizer states. We measure stabilizer states and states generated by a hardware efficient circuit with random parameters which are expected to have a lot of Bell magic (see Appendix~\ref{sec:hardwareefficient}). The measured Bell magic of the states is shown in Appendix~\ref{sec:learndata}. Due to noise, the prepared stabilizer states are mixed states and have non-zero Bell magic. Thus, we use the supervised learning algorithm shown in Appendix~\ref{sec:suplearn} to learn the best decision boundary $\mathcal{B}^*$ from the experimental data. When for a given measured state $\hat{\mathcal{B}}<\mathcal{B}^*$, we say the measured state is a stabilizer state, else we say that it is not a stabilizer state.
In Fig.\ref{fig:ionqlearn}, we show the classification error as a function of the number of samples $N_\text{Q}$ measured on the quantum computer. We find that the experimental results fit well with a simulation of the protocol with the measured depolarizing noise $p=0.15$.  Note that with our machine learning algorithm the maximal error rate is $P_\text{E}^\text{max}=\frac{1}{2}$ due to the trivial strategy of classifying states at random.

\sectionMain{Variational Bell magic solver}\label{sec:magicvariational}

\begin{algorithm}[h]
 \SetAlgoLined
 \LinesNumbered
  \SetKwInOut{Input}{Input}
  \SetKwInOut{Output}{Output}
   \Input{
   
   $j=1,\dots,3N_\text{Q}$ bitstrings $r^j\in\{0,1\}^{2N}$ sampled from Bell measurement on $\ket{\psi(\boldsymbol{\theta})}\otimes\ket{\psi(\boldsymbol{\theta})}$
   
   $j=1,\dots,N_\text{Q}$ bitstrings $q_+^j\in\{0,1\}^{2N}$ sampled from Bell measurement on $\ket{\psi(\boldsymbol{\theta}+\frac{\pi}{2}\boldsymbol{e}_k)}\otimes\ket{\psi(\boldsymbol{\theta})}$
   
   $j=1,\dots,N_\text{Q}$ bitstrings $q_-^j\in\{0,1\}^{2N}$ sampled from Bell measurement on $\ket{\psi(\boldsymbol{\theta}-\frac{\pi}{2}\boldsymbol{e}_k)}\otimes\ket{\psi(\boldsymbol{\theta})}$
   Resampling steps $N_\text{R}$
   }
    \Output{Gradient of Bell magic $\partial_k\mathcal{B}$
    }

$\mathcal{B}_+\leftarrow 0$

$\mathcal{B}_-\leftarrow 0$

 \SetKwRepeat{Do}{do}{while}
    \For{$k=1,\dots,N_\text{R}$}{
    Choose randomly without replacement $n_1,n_2,n_3\in\{1,\dots,3N_\text{Q}\}$
    
    $m\in\{1,\dots,N_\text{Q}\}$
        
    $\mathcal{B}_+\leftarrow \mathcal{B}_+ +\text{Check-Commute}(r^{n_1}\oplus r^{n_2},r^{n_3}\oplus q_+^m)$
    
    $\mathcal{B}_-\leftarrow \mathcal{B}_- +\text{Check-Commute}(r^{n_1}\oplus r^{n_2},r^{n_3}\oplus q_-^m)$
    }
    $\mathcal{B}_+\leftarrow 4\mathcal{B}_+/N_\text{R}$
    
    $\mathcal{B}_-\leftarrow 4\mathcal{B}_-/N_\text{R}$
    
    $\partial_k\mathcal{B}\leftarrow\mathcal{B}_+-\mathcal{B}_-$
 \caption{Gradient of Bell magic}
 \label{alg:bellmagicGrad}
\end{algorithm}

Variational quantum algorithms find the parameters $\boldsymbol{\theta}$ of a parameterized quantum circuit $\ket{\psi(\boldsymbol{\theta})}$ such that they maximize a cost function $C(\boldsymbol{\theta})$ measured on a quantum computer~\cite{peruzzo2014variational,cerezo2020variational,bharti2021noisy}. The algorithm runs in a quantum-classical feedback loop, where the cost function is measured on the quantum computer and used by a classical optimization routine to find improved parameters. 
We now propose a variational quantum algorithm to maximize Bell magic (see Fig.\ref{fig:sketch}d). 
Commonly, the cost function is maximized with gradient descent, where the $k$-th parameter is iteratively updated with the gradient $\boldsymbol{\theta}'_k=\boldsymbol{\theta}_k-\partial_k C(\boldsymbol{\theta})$.  
The shift-rule provides exact gradients when the circuit is composed of parameterized Pauli rotations~\cite{schuld2019evaluating}. For standard measurements on single quantum states, the shift rule is given by $\partial_k\langle C(\boldsymbol{\theta})\rangle=v(\langle C(\boldsymbol{\theta}+\boldsymbol{e}_k\frac{\pi}{4v})\rangle-\langle C(\boldsymbol{\theta}-\boldsymbol{e}_k\frac{\pi}{4v})\rangle)$, where $\boldsymbol{e}_k$ is the $k$th unit vector and $v>0$~\cite{mitarai2018quantum}. 

We extend the shift rule to Bell measurements (see Appendix~\ref{sec:shiftrule})
\begin{align*}
&\partial_k P(\boldsymbol{r})=
2v\bra{\psi(\boldsymbol{\theta}+\frac{\pi}{4v}\boldsymbol{e}_k)}\bra{\psi(\boldsymbol{\theta})} O_{\boldsymbol{r}}\ket{\psi(\boldsymbol{\theta}+\frac{\pi}{4v}\boldsymbol{e}_k)}\ket{\psi(\boldsymbol{\theta})}\\
&-2v\bra{\psi(\boldsymbol{\theta}-\frac{\pi}{4v}\boldsymbol{e}_k)}\bra{\psi(\boldsymbol{\theta})} O_{\boldsymbol{r}}\ket{\psi(\boldsymbol{\theta}-\frac{\pi}{4v}\boldsymbol{e}_k)}\ket{\psi(\boldsymbol{\theta})}\,\numberthis\label{eq:shift_rule}
\end{align*}
and the gradient of Bell magic is given by
\begin{equation}\label{eq:BellMagicGrad}
\partial_k \mathcal{B}=4\sum_{\substack{\boldsymbol{r},\boldsymbol{r'},\boldsymbol{q},\boldsymbol{q'}\\\in\{0,1\}^{2N}}}[\partial_k P(\boldsymbol{r})]P(\boldsymbol{r'})P(\boldsymbol{q})P(\boldsymbol{q'})\left\lVert[\sigma_{\boldsymbol{r}\oplus\boldsymbol{r'}},\sigma_{\boldsymbol{q}\oplus\boldsymbol{q'}}]\right\rVert_{\infty}
\end{equation}
Algorithm~\ref{alg:bellmagicGrad} depicts how to efficiently measure the gradient on a quantum computer for the case $v=\frac{1}{2}$.

Conveniently, the Bell measurements also give us access to the diagonal entries $\mathcal{F}_ {kk}(\boldsymbol{\theta})$ of the quantum Fisher information metric $\mathcal{F}_ {ij}(\boldsymbol{\theta})=4[\braket{\partial_i \psi}{\partial_j \psi}-\braket{\partial_i \psi}{\psi}\braket{\psi}{\partial_j \psi}]$
without requiring additional measurements. The quantum Fisher information metric and its diagonal approximation can tremendously speed up the training of variational quantum algorithms with the quantum natural gradient $\mathcal{F}^{-1}(\boldsymbol{\theta})\nabla C(\boldsymbol{\theta})$~\cite{stokes2020quantum,van2020measurement,haug2021optimal,haug2021natural}. 
With the shift-rule, the diagonal entries of the metric are given by $\mathcal{F}_{kk}(\boldsymbol{\theta})=2(1-\vert\braket{\psi(\boldsymbol{\theta})}{\psi(\boldsymbol{\theta}+\boldsymbol{e}_k\frac{\pi}{2})}\vert^2)$ \cite{mari2021estimating}. For pure states, the fidelity $\vert\braket{\psi(\boldsymbol{\theta})}{\psi(\boldsymbol{\theta}+\boldsymbol{e}_k\frac{\pi}{2})}\vert^2$ is given by the SWAP test \eqref{eq:SWAP}, where we can simply re-use the measurement outcomes for the gradient of magic.

\begin{figure}[htbp]
	\centering	
	\subfigimg[width=0.33\textwidth]{}{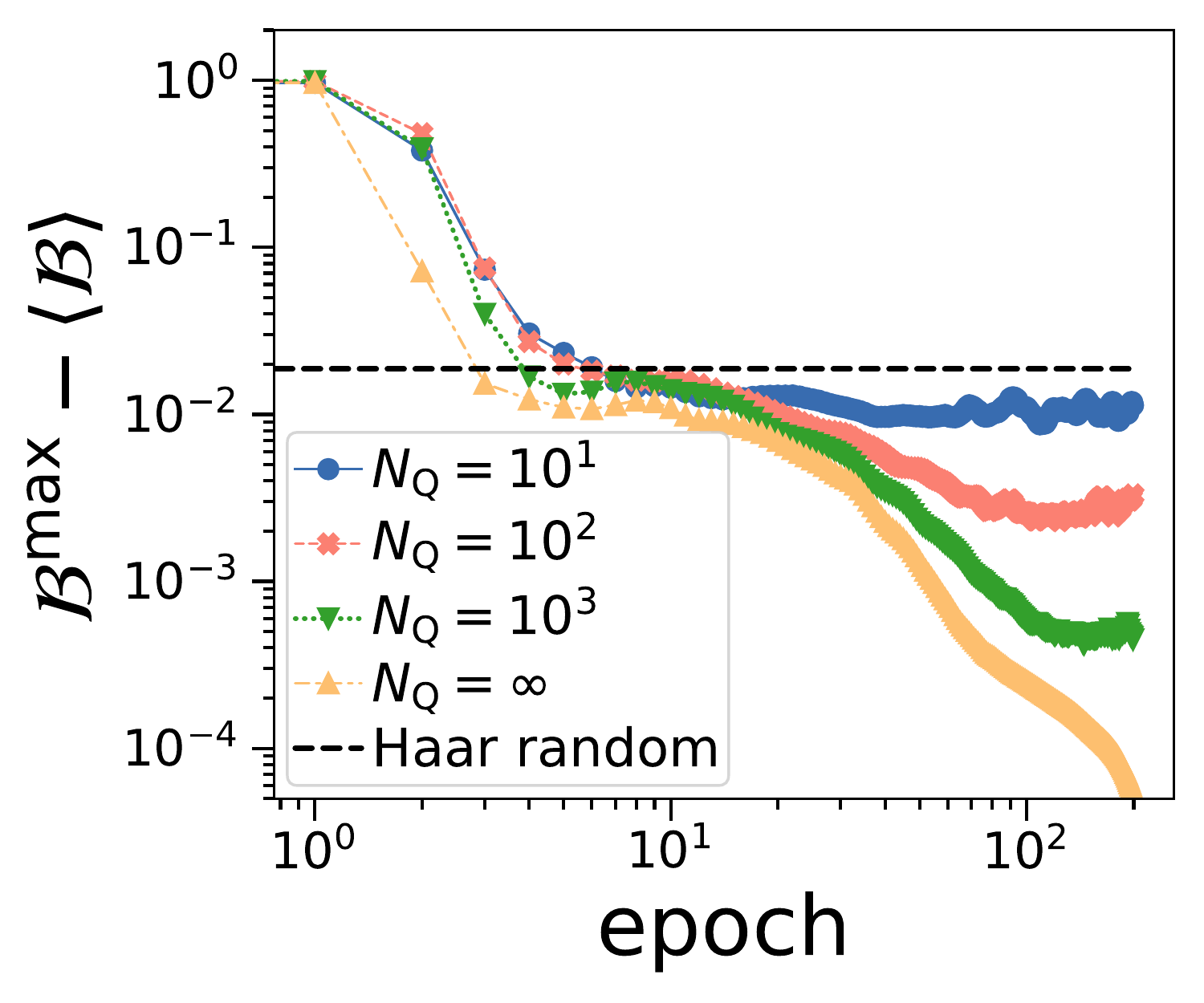}
	\caption{Simulation of variational algorithm to maximize Bell magic $\mathcal{B}$ of a parameterized quantum circuit for varying number of measurement samples $N_\text{Q}$. We plot the difference between the average Bell magic $\langle B \rangle$ found at a training epoch and the maximal possible magic $B^\text{max}$ of pure states. The training results are averaged over 10 random training instances. 
	The learning rate is $\gamma=0.1$, depth of circuit $d=6$ and number of qubits $N=4$.
	}
	\label{fig:train}
\end{figure}

In Fig.\ref{fig:train}, we numerically study the variational Bell magic solver to find the pure state with maximal Bell magic $\mathcal{B}^\text{max}$. We show the variational magic solver to maximize $\mathcal{B}$ as function of training epochs for varying number of measurement samples $N_\text{Q}$. We use the gradient method Adam~\cite{kingma2014adam} and the gradients are determined using the shift-rule, where we use $N_\text{Q}$ measurement samples for each measurement setting. To estimate the gradients, we require in total $N_\text{Q}(2K+3)$ measurement samples, where $K$ is the number of parameters of the circuit. The parameterized quantum circuit is shown in Appendix~\ref{sec:hardwareefficient} and the initial parameters are chosen such that the initial state is close to a stabilizer state. Training is initially fast until reaching the average magic of Haar random quantum states. Then, optimization continues at a slower pace. With increasing $N_\text{Q}$ our solver finds states which have close to the maximal amount of Bell magic.
\revA{In general, we find that pure states of high Bell magic are characterized by having a small, but non-zero expectation value for nearly all Pauli operators. We further study the structure of pure states with high Bell magic in Appendix~\ref{sec:maxmagic}.}

The performance of variational quantum algorithms is tied to their expressibility and trainability~\cite{bharti2021noisy}. 
Expressibility describes how well an ansatz is uniformly explores the full Hilbertspace, which makes it more likely that the ansatz can express the target solution~\cite{sim2019expressibility}. A circuit is maximally expressible if it forms a $2$-design, i.e. averaging over the ansatz matches an average over Haar random unitaries up to the second moment.
A variational quantum algorithms is trainable when the magnitude of gradients is large. A common problem is so-called barren plateau problem, where the magnitude of the gradients vanish exponentially with the number of qubits~\cite{mcclean2018barren}.
Nearly all variational quantum algorithms use the observable $\mathcal{H}=\sum_{n=1}^{\text{poly}(N)}\gamma_n \sigma_{n}$ that consist of a sum over a polynomial number of arbitrary Pauli strings $\sigma_{n}$ with constant coefficients $\gamma_n$. For this general class of cost function, high expressibility directly leads to vanishing gradients and the training becomes impractical~\cite{mcclean2018barren,holmes2021connecting}. 

Bell magic does not belong to the aforementioned class of cost functions as it cannot be expressed by a polynomial sum of Pauli strings and requires two copies of a quantum state to be measured.
We illustrate this fact for a highly expressible ansatz where barren plateaus are absent for Bell magic, while generally used cost functions suffer from barren plateaus. As a simple demonstration, we define the ansatz $\ket{\psi(\theta,U_\text{C})}=U_\text{C}\exp(-i\frac{1}{2}\theta \sigma^y_1)\ket{0}^{\otimes N}$, where $U_\text{C}$ is a Clifford circuit, $\theta$ the parameter of the circuit and $\sigma^y_1$ is the $y$-Pauli operator acting on the first qubit. Over randomly sampled Clifford circuits $U_{\text{C}}$ from the Clifford group $\mathcal{C}$, this ansatz is maximally expressible as it uniformly explores the full Hilbertspace and forms a $2$-design~\cite{webb2015clifford}. Thus, for any cost function $\mathcal{H}$ consisting of a polynomial number of Pauli strings, the magnitude of the gradient decays exponentially with number $N$ of qubits~\cite{mcclean2018barren}
\begin{equation}
    \text{Var}[\partial_{\theta}\bra{\psi(\theta,U_{\text{C})}} \mathcal{H}\ket{\psi(\theta,U_{\text{C}})})]_{\theta,\mathcal{C}}=\frac{\text{Tr}(\mathcal{H}^2)}{2(2^{2N}-1)}\propto 2^{-N}\,,
\end{equation}
where the variance is taken over $\theta$ and the Clifford group $\mathcal{C}$.
In contrast, the variance of the gradient in respect to Bell magic is independent of $N$. As shown in Appendix~\ref{sec:express} we find 
\begin{equation}
    \text{Var}[\partial_{\theta} \mathcal{B}(\ket{\psi(\theta,U_{\text{C}})})]_{\theta,\mathcal{C}}=\frac{1}{2}\,,
\end{equation}
ensuring the trainability of this ansatz for any $N$.

\sectionMain{Discussion}\label{sec:discussion}

We showed how to measure and learn Bell magic with quantum computers. Our algorithm relies on Bell measurements of two copies of a quantum state, which is straightforward to implement on quantum computers and simulators as demonstrated in past experiments to measure entanglement~\cite{islam2015measuring,kaufman2016quantum,huang2021demonstrating}. \revA{Bell magic can be measured concurrently with entanglement, which we verify in Appendix~\ref{sec:entangle} by measuring the Wallach-Meyer entanglement measure.}
For quantum computers, the Bell transformation can be implemented directly without SWAP gates when all qubits are connected to all other qubits such as on the IonQ quantum computer~\cite{wright2019benchmarking}, or when the two copies of the quantum state can be arranged in a ladder structure~\cite{huang2021demonstrating}. Further, it can be easily implemented in numerical simulations to study the magic of many-body quantum systems~\cite{liu2020many,white2021conformal}.
In contrast to existing measures of magic, the measurement cost $N_\text{Q}$ is independent of qubit number and has only an inverse quadratic scaling $N_\text{Q}\propto \Delta\mathcal{B}^{-2}$ with the estimation error $\Delta\mathcal{B}$. 
Noise occurring in experiments can be mitigated with a scaling of $N_\text{Q}\propto(1-p)^{-16}$. We derive these bounds assuming a low number of resampling steps $N_\text{R}=N_\text{Q}/4$ in the classical part of the algorithm, where a better scaling is possible by increasing $N_\text{R}$. 

For our study we assume depolarizing noise, which we find is sufficient to mitigate noise on the IonQ quantum computer. The error mitigated Bell magic of our experiments closely matches exact simulations, opening the study of non-stabilizerness on noisy quantum computers~\cite{bharti2021noisy}. 
Additional methods that turn non-depolarizing noise into depolarizing noise could be used to further improve error mitigation~\cite{wallman2016noise,urbanek2021mitigating}.

Fault-tolerant quantum computers are commonly run by state synthesis protocols, where a resource state $\ket{\phi}_\text{ini}$ combined with error-corrected Clifford operations is transformed into a target state $\ket{\phi}_\text{target}$~\cite{campbell2017roads}. Magic quantifies \revA{a lower bound} on the non-stabilizer resources needed to synthesize a particular state or unitary~\cite{howard2017application,beverland2020lower,bravyi2016trading}. As Bell magic is invariant under Clifford unitaries, a necessary condition for state synthesis is that $\mathcal{B}(\ket{\phi}_\text{target})\le\mathcal{B}(\ket{\phi}_\text{ini})$. Thus, Bell magic can experimentally establish lower bounds on the magical resources needed to synthesize states on quantum computers.

\revA{Another important task is verifying whether a given state is indeed correct~\cite{eisert2020quantum,carrasco2021theoretical}. Given a stabilizer state, learning the description of the state requires measurements on $O(N)$ copies~\cite{montanaro2017learning}. 
In contrast, discriminating whether a given state is a stabilizer state is comparatively easier, requiring only $O(1)$ copies~\cite{gross2021schur}.
Our work allows to distinguish states with different degree of non-stabilizerness in the presence of noise.} For example, with only $N_\text{Q}=6$ measurements, we can decide with an error of less than 1\% whether a given state is a stabilizer state or a highly magical state. To distinguish stabilizer states and near-stabilizer states as defined in \eqref{eq:low_magic_state}, we find a scaling of $N_\text{Q}\propto\phi^{-2}$. 
Our method could be used to reliably certify states for quantum communication~\cite{gisin2007quantum} and quantum computing tasks~\cite{gottesman1998theory}. Note that the related task of testing whether a given unitary is Clifford has been studied in~\cite{low2009learning,gross2021schur}.

We also find that Bell magic is connected to 
the recently proposed linear stabilizer entropy $M_\text{lin}=1-2^{N}\sum_{\boldsymbol{r}}(2^{-N}\bra{\psi}\sigma_{\boldsymbol{r}}\ket{\psi}^2)^2$
and $2$-Rényi entropy $M_2=-\log(2^{N}\sum_{\boldsymbol{r}}(2^{-N}\bra{\psi}\sigma_{\boldsymbol{r}}\ket{\psi}^2)^2)$~\cite{leone2021renyi}. 
We find that these measures of magic can be also computed with Bell measurements via $M_\text{lin}=1-2^{N}\sum_{\boldsymbol{r}}P(\boldsymbol{r})^2$ and $M_2=-\log(2^{N}\sum_{\boldsymbol{r}}P(\boldsymbol{r})^2)$ (see Appendix~\ref{sec:connect_stabilizer_sup}). Here, one has to explicitly estimate the probabilities $P(\boldsymbol{r})$ which in general requires an exponential amount of measurements. We experimentally compute stabilizer entropy with the IonQ quantum computer and compare it with Bell magic in the Appendix~\ref{sec:connect_stabilizer_sup}, where we find that both measures behave similarly.

Our results reveal a fundamental relationship between magic and the probability distribution of bitstrings in the Bell basis. Statistical tests over these distributions could be performed, similar to the cross entropy benchmarking in the computational basis for quantum supremacy~\cite{arute2019quantum}. As magic is closely related to the classical complexity of simulating quantum states, it could serve as an alternative test for quantum supremacy.

\revA{Bell magic is a faithful measure of magic for pure states, while mixed Bell magic is also faithful for a class of mixed states~\eqref{eq:mixedmagic}.
We proved the invariance under Clifford unitaries and composition, and our numerics suggest that Bell magic is non-increasing under partial trace and measurements in the computational basis. We leave the formal proof whether mixed Bell magic fulfills these conditions as an open problem~\cite{veitch2014resource}.
Using convex roof type extensions it may also be possible to construct a faithful Bell magic for arbitrary mixed states~\cite{seshadreesan2015renyi}. An extension to unitaries, channel capacity and qudits would interesting as well~\cite{veitch2014resource,wang2019quantifying}. It would be also useful to find a quantitative connection between Bell magic and classical simulation complexity~\cite{pashayan2015estimating}.}

Finally, we showed that our variational algorithm for Bell magic can have large gradients even for highly expressible circuits that uniformly sample the full Hilbertspace. For commonly used cost functions, expressible circuits must have barren plateaus~\cite{mcclean2018barren,cerezo2021cost,holmes2021connecting,marrero2021entanglement}. However, this rule does not apply to Bell magic as it cannot be expressed by a polynomial number of Pauli strings. As example, we demonstrate an ansatz which has a high expressibility yet the gradient is independent of $N$. 
Unbounded cost functions are known to have similar features, although they are not suited for near-term quantum computers~\cite{kieferova2021quantum}. We note that barren plateaus can still appear depending on the choice of ansatz.
It would be interesting to search for other cost functions that combine expressibility and large gradients by using entangled measurements over multiple copies~\cite{chen2021exponential}. 
One could also study the trainability of Bell magic in conjunction with the learnability transition that occurs in Clifford circuits combined with T-gates~\cite{zhou2020single,true2022transitions}.

The code for this paper is available at~\cite{haug2022bellmagic}.

\textit{Note added:} While finishing the manuscript, the stabilizer entropy was experimentally measured with a randomized measurement protocol that scales exponentially with the number of qubits~\cite{oliviero2022measuring}.

 \let\oldaddcontentsline\addcontentsline
\renewcommand{\addcontentsline}[3]{}

\medskip
\begin{acknowledgments}
We thank Kishor Bharti, Ronan Docherty, Nikolaos Koukoulekidis, Ashley Montanaro, Felix Roberts, David Steuerman and Mark Wilde for insightful comments. We thank IonQ for providing quantum computing resources.
This work is supported by a Samsung GRC project and the UK Hub in Quantum Computing and Simulation, part of the UK National Quantum Technologies Programme with funding from UKRI EPSRC grant EP/T001062/1. 
\end{acknowledgments}
\bibliography{learnmagic}

\let\addcontentsline\oldaddcontentsline

\clearpage
\appendix

\tableofcontents

\section{Stabilizer states and Bell measurement}\label{sec:Stabcommute}

Here, we show the connection between Bell measurement and stabilizer states.
The transformation into the Bell basis is applied with the unitary $U_\text{Bell}=\bigotimes_{n=1}^N (H\otimes I_2)\cdot\text{CNOT}$ on all $N$ qubit pairs (see Fig.\ref{fig:sketch}b). Here, $H$ is the Hadamard gate and $\text{CNOT}$ denotes the CNOT gate. The resulting state is sampled $N_\text{Q}$ times and record the $j$th measurement outcome as bitstring
$\boldsymbol{r}^j\in\{0,1\}^{2N}$. 
The outcome $\boldsymbol{r}$ appears with a probability~\cite{montanaro2017learning}
\begin{equation}\label{eq:probBell_sup}
P(\boldsymbol{r})=\bra{\psi}\bra{\psi}O_{\boldsymbol{r}}\ket{\psi}\ket{\psi}=2^{-N}\vert\bra{\psi}\sigma_{\boldsymbol{r}}\ket{\psi^*}\vert^2\,,
\end{equation}
where $O_{\boldsymbol{r}}=\ket{\sigma_{\boldsymbol{r}}}\bra{\sigma_{\boldsymbol{r}}}$ is the projector onto the Bell state and $\ket{\psi^*}$ denotes the complex conjugate of $\ket{\psi}$. For any state, there are at least $2^N$ possible outcomes $\boldsymbol{r}$ as $\vert\bra{\psi}\sigma_{\boldsymbol{r}}\ket{\psi^*}\vert^2\le1$.

Stabilizer states $\ket{\psi_\text{STAB}}$ are defined by a commuting subgroup $G$ of $\abs{G}=2^N$ Pauli strings $\sigma$. We have $\bra{\psi_\text{STAB}}\sigma\ket{\psi_\text{STAB}}=\pm1$ for $\sigma\in G$ and  $\bra{\psi_\text{STAB}}\sigma'\ket{\psi_\text{STAB}}=0$ for $\sigma'\notin G$. Any $\sigma_{\boldsymbol{r}},\sigma_{\boldsymbol{r}'}\in G$ commute $[\sigma_{\boldsymbol{r}},\sigma_{\boldsymbol{r}'}]=0$.
Any stabilizer state can be written as~\cite{dehaene2003clifford}
\begin{equation}
 \ket{\psi_\text{STAB}} = \frac{1}{\sqrt{|A|}} \sum_{x \in A} i^{\ell(x)} (-1)^{q(x)} \ket{x}\,, 
 \end{equation}
where $A$ is an affine subspace of the Galois field $\mathbb{F}_2^N$ and $\ell,z: \{0,1\}^N \rightarrow \{0,1\}$ are linear and quadratic polynomials over $\mathbb{F}_2$. As $\ell$ is linear, we have $\ell(x) = s \cdot x$ for $s \in \{0,1\}^N$ and the complex factor of the Stabilizer state can be written as $i^{\ell(x)} = \prod_{k \in S} i^{x_k}$ for some $S \subseteq [N]$. 
Thus, we can write the complex conjugate as a transformation with the $z$ Pauli operator
$\ket{\psi^*} = \sigma_{10}^{\otimes S}\ket{\psi}=\sigma_{\boldsymbol{g}}\ket{\psi}$, with some $\boldsymbol{g}=\{s_1,0,s_2,0,\dots,s_N,0\}$ that characterizes the complex phase of the stabilizer state~\cite{montanaro2017learning}.
Inserting this relation into \eqref{eq:probBell_sup}, the probability of sampling a bitstring $\boldsymbol{q}$ from a stabilizer state is given as $P(\boldsymbol{q})=2^{-N}\vert\bra{\psi_\text{STAB}}\sigma_{\boldsymbol{q}}\sigma_{\boldsymbol{g}}\ket{\psi_\text{STAB}}\vert^2=2^{-N}\vert\bra{\psi_\text{STAB}}\sigma_{\boldsymbol{q}\oplus\boldsymbol{g} }\ket{\psi_\text{STAB}}\vert^2$. There are $2^N$ outcomes with non-zero probability $P(\boldsymbol{q})>0$  with $\sigma_{\boldsymbol{t}}=\sigma_{\boldsymbol{q}\oplus\boldsymbol{g} }\in G$. Any outcome $\boldsymbol{q}$ can be written as $\boldsymbol{q}=\boldsymbol{t}\oplus\boldsymbol{g}$ where the set of strings $\boldsymbol{t} \in \{0,1\}^{2N}$ with $\sigma_{\boldsymbol{t}} \in G$ forms an $N$-dimensional linear subspace of $\mathbb{F}_2^{2N}$. The addition of two elements of the subspace $\boldsymbol{t}\oplus\boldsymbol{t}'$ is again part of the commuting subspace $\sigma_{\boldsymbol{t}\oplus\boldsymbol{t}'} \in G$.
The addition of two measured outcomes $\boldsymbol{q},\boldsymbol{q}'$ yields $\boldsymbol{q}\oplus\boldsymbol{q}'=\boldsymbol{t}\oplus\boldsymbol{g}\oplus\boldsymbol{t}'\oplus\boldsymbol{g}=\boldsymbol{t}\oplus\boldsymbol{t}'$. Thus, the addition of two outcomes from a stabilizer state yields a Pauli string of the commuting subgroup $\sigma_{\boldsymbol{q}\oplus\boldsymbol{q}'} \in G$.

In summary, for any set of bitstrings $\{\boldsymbol{q}_n\}_{n=1}^{N_\text{Q}}$ sampled in the Bell basis from a pure stabilizer state, the Pauli strings of its binary additions must commute, i.e. $[\sigma_{\boldsymbol{q}_k\oplus\boldsymbol{q}_l},\sigma_{\boldsymbol{q}_n\oplus\boldsymbol{q}_m}]=0$ $\forall k,l,n,m$. Conversely, finding at least one non-commuting Pauli string implies that the measured quantum state is not a pure stabilizer state as the commuting subgroup $G$ contains at most $2^N$ elements.

\section{Invariance under Clifford circuits}\label{sec:invariance}
We now show that Bell magic is invariant $\mathcal{B}(\ket{\psi})=\mathcal{B}(U_\text{C}\ket{\psi})$ under arbitrary Clifford circuits $U_\text{C}$, i.e. unitaries that map stabilizer states into stabilizer states.

To this end, we start by proving that the second and third line of \eqref{eq:to_show} are indeed equal
\begin{align*}
&Q(\boldsymbol{n})=\sum_{\boldsymbol{r}\in\{0,1\}^{2N}}P(\boldsymbol{r})P(\boldsymbol{r}\oplus\boldsymbol{n})\equiv\\
&\sum_{\boldsymbol{r}}\bra{\psi}\bra{\psi}O_{\boldsymbol{r}}\ket{\psi}\ket{\psi} \bra{\psi}\bra{\psi}O_{\boldsymbol{r}\oplus\boldsymbol{n}}\ket{\psi} \ket{\psi}=\\
&4^{-N}\sum_{\boldsymbol{r}}\bra{\psi}\sigma_{\boldsymbol{r}}\ket{\psi}^2\bra{\psi}\sigma_{\boldsymbol{r}\oplus\boldsymbol{n}}\ket{\psi}^2\numberthis\label{eq:to_show}\,,
\end{align*}
where we have an arbitrary $\boldsymbol{n}\in\{0,1\}^{2N}$.
We note that by using 4 copies of $\ket{\psi}$, we can rewrite the bottom line of \eqref{eq:to_show} as 
\begin{equation}\label{eq:tensor_Pauli}
\sum_{\boldsymbol{r}}\bra{\psi}\sigma_{\boldsymbol{r}}\ket{\psi}^2\bra{\psi}\sigma_{\boldsymbol{r}\oplus\boldsymbol{n}}\ket{\psi}^2=\bra{\psi}^{\otimes 4}\sum_{\boldsymbol{r}}\sigma_{\boldsymbol{r}}^{\otimes 2}\otimes \sigma_{\boldsymbol{r}\oplus\boldsymbol{n}}^{\otimes 2}\ket{\psi}^{\otimes 4}.
\end{equation}
Similarly, the second line of \eqref{eq:to_show} can be rewritten  as 
\begin{align*}
&\sum_{\boldsymbol{r}}\bra{\psi}\bra{\psi}O_{\boldsymbol{r}}\ket{\psi}\ket{\psi} \bra{\psi}\bra{\psi}O_{\boldsymbol{r}\oplus\boldsymbol{n}}\ket{\psi} \ket{\psi}=\\
&\bra{\psi}^{\otimes 4}\sum_{\boldsymbol{r}}O_{\boldsymbol{r}}\otimes O_{\boldsymbol{r}\oplus\boldsymbol{n}} \ket{\psi}^{\otimes 4}\numberthis\label{eq:tensor_Bell}
\end{align*}
Now, we note that both the Bell operator $O_{\boldsymbol{r}}$ and the Pauli string $\sigma_{\boldsymbol{r}}$ can be written as tensor products acting on 2 and 1 qubit respectively.
In particular,
\begin{equation}
    O_{\boldsymbol{r}}=\ket{\sigma_{r_1r_2}}\bra{\sigma_{r_1r_2}}\otimes\dots\otimes\ket{\sigma_{r_{2N-1}r_{2N}}}\bra{\sigma_{r_{2N-1}r_{2N}}}
\end{equation}
and
\begin{equation}
\sigma_{\boldsymbol{r}}=\sigma_{r_1r_2}\otimes\dots\otimes\sigma_{r_{2N-1}r_{2N}}
\end{equation}
At the same time, the sum can be decomposed as $\sum_{\boldsymbol{r}}=\sum_{r_1}\sum_{r_2}\dots\sum_{r_{2N}}$.

The operators can be written as tensor products of $N$ operators acting on 4-qubit subspaces
\begin{align*}
&\sum_{\boldsymbol{r}}\sigma_{\boldsymbol{r}}^{\otimes 2}\otimes \sigma_{\boldsymbol{r}\oplus\boldsymbol{n}}^{\otimes 2}=\sum_{r_1,r_2}\sigma_{r_1 r_2}^{\otimes 2}\otimes\sigma_{r_1\oplus n_1 r_2\oplus n_2}^{\otimes 2}\otimes\dots\otimes\\
&\sum_{r_{2N-1}r_{2N}}\sigma_{r_{2N-1}r_{2N}}^{\otimes 2}\otimes\sigma_{r_{2N-1}\oplus n_{2N-1} r_{2N}\oplus n_{2N}}^{\otimes 2}
\end{align*}
and
\begin{align*}
&\sum_{\boldsymbol{r}}O_{\boldsymbol{r}}\otimes O_{\boldsymbol{r}\oplus\boldsymbol{n}}=\\
&\sum_{r_1,r_2}\ket{\sigma_{r_1r_2}}\bra{\sigma_{r_1r_2}}\otimes\ket{\sigma_{r_1\oplus n_1 r_2\oplus n_2}}\bra{\sigma_{r_1\oplus n_1 r_2\oplus n_2}}
\otimes\dots\otimes\\
&\sum_{r_{2N-1}r_{2N}}\ket{\sigma_{r_{2N-1}r_{2N}}}
\bra{\sigma_{r_{2N-1}r_{2N}}}\otimes\\
&\ket{\sigma_{r_{2N-1}\oplus n_{2N-1} r_{2N}\oplus n_{2N}}}\bra{\sigma_{r_{2N-1}\oplus n_{2N-1} r_{2N}\oplus n_{2N}}}
\end{align*}

Now, one can explicitly calculate the operators on the 4-qubit Hilbertspace by hand and we find for all $n_1,n_2\in\{0,1\}$ the following equivalence
\begin{align*}
&\sum_{r_1,r_2}\ket{\sigma_{r_1r_2}}\bra{\sigma_{r_1r_2}}\otimes\ket{\sigma_{r_1\oplus n_1 r_2\oplus n_2}}\bra{\sigma_{r_1\oplus n_1 r_2\oplus n_2}}=\\
&\frac{1}{4}\sum_{r_1,r_2}\sigma_{r_1 r_2}^{\otimes 2}\otimes\sigma_{r_1\oplus n_1 r_2\oplus n_2}^{\otimes 2}
\end{align*}
This equivalence also holds for any tensor product of the 4-qubit operator. 
Thus, we find for any $\boldsymbol{n}\in\{0,1\}^{2N}$
\begin{equation}
\sum_{\boldsymbol{r}}O_{\boldsymbol{r}}\otimes O_{\boldsymbol{r}\oplus\boldsymbol{n}}=\frac{1}{4^N}\sum_{\boldsymbol{r}}\sigma_{\boldsymbol{r}}^{\otimes 2}\otimes \sigma_{\boldsymbol{r}\oplus\boldsymbol{n}}^{\otimes 2}\,.
\end{equation}
Together with \eqref{eq:tensor_Pauli} and \eqref{eq:tensor_Bell}, this proves that \eqref{eq:to_show} is indeed correct. 

As a corollary, combining \eqref{eq:to_show} and \eqref{eq:probBell_sup} yields the surprising relation
\begin{align*}
&\sum_{\boldsymbol{r}\in\{0,1\}^{2N}}\bra{\psi}\sigma_{\boldsymbol{r}}\ket{\psi}^2\bra{\psi}\sigma_{\boldsymbol{r}\oplus\boldsymbol{n}}\ket{\psi}^2=\\
&\sum_{\boldsymbol{r}\in\{0,1\}^{2N}}\vert\bra{\psi}\sigma_{\boldsymbol{r}}\ket{\psi^*}\vert^2\vert\bra{\psi}\sigma_{\boldsymbol{r\oplus\boldsymbol{n}}}\ket{\psi^*}\vert^2
\numberthis\label{eq:result_conj}\,.
\end{align*}

Next, we proceed to prove the invariance of Bell magic.
As reminder, Bell magic is defined as
\begin{equation}
\mathcal{B}=\sum_{\substack{\boldsymbol{r},\boldsymbol{r'},\boldsymbol{q},\boldsymbol{q'}\\\in\{0,1\}^{2N}}}P(\boldsymbol{r})P(\boldsymbol{r'})P(\boldsymbol{q})P(\boldsymbol{q'})\left\lVert[\sigma_{\boldsymbol{r}\oplus\boldsymbol{r'}},\sigma_{\boldsymbol{q}\oplus\boldsymbol{q'}}]\right\rVert_{\infty}
\end{equation}
Also, remember that $\sigma_{\boldsymbol{r}\oplus\boldsymbol{r'}}=\sigma_{\boldsymbol{r}}\sigma_{\boldsymbol{r'}}$ up to a prefactor $\{1,-1,i,-i\}$.
Now, we equivalently rewrite Bell magic into
\begin{equation}
\mathcal{B}=\sum_{\boldsymbol{n},\boldsymbol{q}\in\{0,1\}^{2N}}Q(\boldsymbol{n})Q(\boldsymbol{q})\left\lVert[\sigma_{\boldsymbol{n}},\sigma_{\boldsymbol{q}}]\right\rVert_{\infty}\,.
\end{equation}
where we define 
$Q(\boldsymbol{n})=\sum_{\boldsymbol{r}}P(\boldsymbol{r})P(\boldsymbol{r}\oplus\boldsymbol{n})$, with $P(\boldsymbol{r})$ as defined in \eqref{eq:probBell_sup}.
Now, we transform $\ket{\psi}$ with a random Clifford circuit $U_\text{C}$ into $U_\text{C}\ket{\psi}$. Note that $U_\text{C}$ transforms a Pauli string $\sigma_{\boldsymbol{n}}$ into another Pauli string $\sigma_{\boldsymbol{q}}$ with $U_\text{C}\sigma_{\boldsymbol{n}} U_\text{C}^\dagger=\sigma_{\boldsymbol{q}}$. This transformation is bijective, i.e. each Pauli string is mapped to another unique Pauli string.
Using \eqref{eq:to_show}, we have $Q(\boldsymbol{n})=4^{-N}\sum_{\boldsymbol{r}}\bra{\psi}\sigma_{\boldsymbol{r}}\ket{\psi}^2\bra{\psi}\sigma_{\boldsymbol{r}}\sigma_{\boldsymbol{n}}\ket{\psi}^2$. The transformed probability $Q'(\boldsymbol{n})$ is given by
\begin{align*}
&Q'(\boldsymbol{n})=4^{-N}\sum_{\boldsymbol{r}}\bra{\psi}U_\text{C}^\dagger\sigma_{\boldsymbol{r}}U_\text{C}\ket{\psi}^2\bra{\psi}U_\text{C}^\dagger\sigma_{\boldsymbol{r}}\sigma_{\boldsymbol{n}}U_\text{C}\ket{\psi}^2=\\
&4^{-N}\sum_{\boldsymbol{r}}\bra{\psi}U_\text{C}^\dagger\sigma_{\boldsymbol{r}}U_\text{C}\ket{\psi}^2\bra{\psi}U_\text{C}^\dagger\sigma_{\boldsymbol{r}}U_\text{C}U_\text{C}^\dagger\sigma_{\boldsymbol{n}}U_\text{C}\ket{\psi}^2=\\
&4^{-N}\sum_{\boldsymbol{r}}\bra{\psi}\sigma_{\boldsymbol{r}}\ket{\psi}^2\bra{\psi}\sigma_{\boldsymbol{r}}\sigma_{\boldsymbol{m}}\ket{\psi}^2\equiv Q(\boldsymbol{m})
\end{align*}
where we used that the sum over all Pauli strings remains invariant due to bijective property and we defined $\boldsymbol{m}$ as the transformed Pauli string $\sigma_{\boldsymbol{m}}=U_\text{C}^\dagger\sigma_{\boldsymbol{n}} U_\text{C}$. This means that a transformation with $U_\text{C}$ simply permutes the distribution $Q(\boldsymbol{n})$.
The Bell magic $\mathcal{B}'$ after transformation is given by 
\begin{align*}
&\mathcal{B}'=\sum_{\boldsymbol{n},\boldsymbol{q}\in\{0,1\}^{2N}}Q'(\boldsymbol{n})Q'(\boldsymbol{q})\left\lVert[\sigma_{\boldsymbol{n}},\sigma_{\boldsymbol{q}}]\right\rVert_{\infty}=\\
&\sum_{\boldsymbol{n},\boldsymbol{q}\in\{0,1\}^{2N}}Q(\boldsymbol{n})Q(\boldsymbol{q})\left\lVert[U_\text{C}\sigma_{\boldsymbol{n}}U_\text{C}^\dagger,U_\text{C}\sigma_{\boldsymbol{q}}U_\text{C}^\dagger]\right\rVert_{\infty}=\\
&\sum_{\boldsymbol{n},\boldsymbol{q}\in\{0,1\}^{2N}}Q(\boldsymbol{n})Q(\boldsymbol{q})\left\lVert U_\text{C}[\sigma_{\boldsymbol{n}},\sigma_{\boldsymbol{q}}]U_\text{C}^\dagger\right\rVert_{\infty}=\\
&\sum_{\boldsymbol{n},\boldsymbol{q}\in\{0,1\}^{2N}}Q(\boldsymbol{n})Q(\boldsymbol{q})\left\lVert[\sigma_{\boldsymbol{n}},\sigma_{\boldsymbol{q}}]\right\rVert_{\infty}
\equiv \mathcal{B}
\end{align*}
where in the final step we used that the commutator of two Pauli strings is either 0 or another Pauli string, and therefore the norm is left invariant under transformation with $U_\text{C}$.

\section{Additive Bell magic}\label{sec:proofadditive}
We now show that the additive Bell magic $\mathcal{B}_\text{a}=-\log_2(1-\mathcal{B})$ is additive. 
We regard a product $\ket{\psi}=\ket{A}\otimes\ket{B}$. of two arbitrary states $\ket{A}$, $\ket{B}$.
We now want to show that $\mathcal{B}_\text{a}(\ket{A}\otimes\ket{B})=\mathcal{B}_\text{a}(\ket{A})+\mathcal{B}_\text{a}(\ket{B})$.

First, we make some preliminary considerations.

As the Bell transformation is a tensor product, the outcomes appearing on the qubits of $\ket{A}$ and $\ket{B}$ are independent of each other. We define the outcomes for the qubits of $\ket{A}$ as $\boldsymbol{r}_A$ and for $\ket{B}$ as $\boldsymbol{r}_B$.

The product of two Pauli strings $\sigma_{\boldsymbol{r}},\sigma_{\boldsymbol{q}}$ is given by $\sigma_{\boldsymbol{r}}\sigma_{\boldsymbol{q}}=\pm i^{C_{\boldsymbol{r},\boldsymbol{q}}}\sigma_{\boldsymbol{r}\oplus\boldsymbol{q}}$, where $C_{\boldsymbol{r},\boldsymbol{q}}=0$ when $[\sigma_{\boldsymbol{r}},\sigma_{\boldsymbol{q}}]=0$ and $C_{\boldsymbol{r},\boldsymbol{q}}=1$ otherwise.
We can use this to write the commutator for the tensor product as
\begin{align*}
&[\sigma_{\boldsymbol{r}},\sigma_{\boldsymbol{q}}]=
[\sigma_{\boldsymbol{r}_A}\otimes\sigma_{\boldsymbol{r}_B},\sigma_{\boldsymbol{q}_A}\otimes\sigma_{\boldsymbol{q}_B}]=\\
&(\sigma_{\boldsymbol{r}_A}\sigma_{\boldsymbol{q}_A})\otimes(\sigma_{\boldsymbol{r}_B}\sigma_{\boldsymbol{q}_B})-(\sigma_{\boldsymbol{q}_A}\sigma_{\boldsymbol{r}_A})\otimes(\sigma_{\boldsymbol{q}_B}\sigma_{\boldsymbol{r}_B})=\\
&\pm\sigma_{\boldsymbol{r}_A\oplus\boldsymbol{q}_A}\otimes\sigma_{\boldsymbol{r}_B\oplus\boldsymbol{q}_B}((-1)^{C_{\boldsymbol{r}_A,\boldsymbol{q}_A}+C_{\boldsymbol{r}_B,\boldsymbol{q}_B}}-\\
&(-1)^{C_{\boldsymbol{r}_A,\boldsymbol{q}_A}+C_{\boldsymbol{r}_B,\boldsymbol{q}_B}})
\end{align*}
This implies that the commutator is non-zero only when $C_{\boldsymbol{r}_A,\boldsymbol{q}_A}+C_{\boldsymbol{r}_B,\boldsymbol{q}_B}=1$, i.e. when we have $[\sigma_{\boldsymbol{r}_A},\sigma_{\boldsymbol{q}_A}]\ne0$ and $[\sigma_{\boldsymbol{r}_B},\sigma_{\boldsymbol{q}_B}]=0$, or $[\sigma_{\boldsymbol{r}_A},\sigma_{\boldsymbol{q}_A}]=0$ and $[\sigma_{\boldsymbol{r}_B},\sigma_{\boldsymbol{q}_B}]\ne0$.  This is the case when the Pauli strings of $A$ commute, but not of $B$, as well as the reverse case. 
With this result, we can now write the infinity norm of the commutator as 
$\left\lVert[\sigma_{\boldsymbol{r}},\sigma_{\boldsymbol{q}}]\right\rVert_{\infty}=\left\lVert[\sigma_{\boldsymbol{r}_A},\sigma_{\boldsymbol{q}_A}]\right\rVert_{\infty}+\left\lVert[\sigma_{\boldsymbol{r}_B},\sigma_{\boldsymbol{q}_B}]\right\rVert_{\infty}-\left\lVert[\sigma_{\boldsymbol{r}_A},\sigma_{\boldsymbol{q}_A}]\right\rVert_{\infty}\left\lVert[\sigma_{\boldsymbol{r}_B},\sigma_{\boldsymbol{q}_B}]\right\rVert_{\infty}$. This expression is zero only when both $A$ and $B$ do not commute, or both commute. Remember that $\left\lVert [\sigma_{\boldsymbol{r}},\sigma_{\boldsymbol{q}}]\right\rVert_{\infty}=2$ when $\sigma_{\boldsymbol{r}},\sigma_{\boldsymbol{q}}$ do not commute, and zero otherwise.

To simplify the notation for the Bell magic, we define
$Q(\boldsymbol{r})=\sum_{\boldsymbol{q}}P(\boldsymbol{q})P(\boldsymbol{q}\oplus\boldsymbol{r})$. Note that $\sigma_{\boldsymbol{q}}\sigma_{\boldsymbol{r}}=\sigma_{\boldsymbol{q}\oplus\boldsymbol{r}}$ up to a multiplication with $\{1,-1,i,-i\}$, where we can ignore this factor for the calculation of the Bell magic since we take the norm of the commutator.
With above fact, we can write the Bell magic as
\begin{equation}
\mathcal{B}=\sum_{\boldsymbol{r},\boldsymbol{q}\in\{0,1\}^{2N}}Q(\boldsymbol{r})Q(\boldsymbol{q})\left\lVert[\sigma_{\boldsymbol{r}},\sigma_{\boldsymbol{q}}]\right\rVert_{\infty}\,.
\end{equation}
As the Bell transformation is a tensor product and the underlying state is a product state, the probability for the outcomes are independent and we can write 
\begin{equation}
    Q(\boldsymbol{r})=Q_A(\boldsymbol{r}_A)Q_B(\boldsymbol{r}_B)\,.
\end{equation}
We note that $\sum_{\boldsymbol{r}_A,\boldsymbol{q}_A}Q_A(\boldsymbol{r}_A)Q_A(\boldsymbol{q}_A)=1$ and $\sum_{\boldsymbol{r}_B,\boldsymbol{q}_B}Q_B(\boldsymbol{r}_B)Q_B(\boldsymbol{q}_B)=1$.

We now combine our considerations to calculate the magic of the product state. We find
\begin{align*}
\mathcal{B}&(\ket{A}\otimes\ket{B})=\\
&\sum_{\boldsymbol{r}_A,\boldsymbol{q}_A}\sum_{\boldsymbol{r}_B,\boldsymbol{q}_B} Q_A(\boldsymbol{r}_A)Q_A(\boldsymbol{q}_A)Q_B(\boldsymbol{r}_B)Q_B(\boldsymbol{q}_B))\\
&\left\lVert[\sigma_{\boldsymbol{r}_A}\otimes\sigma_{\boldsymbol{r}_B},\sigma_{\boldsymbol{q}_A}\otimes\sigma_{\boldsymbol{q}_B}]\right\rVert_{\infty}=\\
&\sum_{\boldsymbol{r}_A,\boldsymbol{q}_A} Q_A(\boldsymbol{r}_A)Q_A(\boldsymbol{q}_A)\left\lVert[\sigma_{\boldsymbol{r}_A},\sigma_{\boldsymbol{q}_A}]\right\rVert_{\infty}+\\
&\sum_{\boldsymbol{r}_B,\boldsymbol{q}_B} Q_B(\boldsymbol{r}_B)Q_B(\boldsymbol{q}_B)\left\lVert[\sigma_{\boldsymbol{r}_B},\sigma_{\boldsymbol{q}_B}]\right\rVert_{\infty}-\\
&(\sum_{\boldsymbol{r}_A,\boldsymbol{q}_A} Q_A(\boldsymbol{r}_A)Q_A(\boldsymbol{q}_A)\left\lVert[\sigma_{\boldsymbol{r}_A},\sigma_{\boldsymbol{q}_A}]\right\rVert_{\infty})\cdot\\
&(\sum_{\boldsymbol{r}_B,\boldsymbol{q}_B} Q_B(\boldsymbol{r}_B)Q_B(\boldsymbol{q}_B)\left\lVert[\sigma_{\boldsymbol{r}_B},\sigma_{\boldsymbol{q}_B}]\right\rVert_{\infty})=\\
&\mathcal{B}(\ket{A})+\mathcal{B}(\ket{B})-\mathcal{B}(\ket{A})\mathcal{B}(\ket{B})
\end{align*}
We now find for the additive magic
\begin{equation}
\mathcal{B}_\text{a}(\ket{A}\otimes\ket{B})=-\log_2[1-\mathcal{B}(\ket{A})-\mathcal{B}(\ket{B})+\mathcal{B}(\ket{A})\mathcal{B}(\ket{B})]
\end{equation}
Finally, the additive magic of the individual states $\ket{A}$ and $\ket{B}$ is given by
\begin{align*}
&\mathcal{B}_\text{a}(\ket{A})+\mathcal{B}_\text{a}(\ket{B})=-\log_2[(1-\mathcal{B}(\ket{A}))(1-\mathcal{B}(\ket{B}))]=\\
&-\log_2[1-\mathcal{B}(\ket{A})-\mathcal{B}(\ket{B})+\mathcal{B}(\ket{A})\mathcal{B}(\ket{B})]\equiv\\
&\mathcal{B}_\text{a}(\ket{A}\otimes\ket{B})\,,
\end{align*}
which concludes our proof.

\section{Composition of Bell magic and stabilizer states}\label{sec:composition}
\revA{We now show that Bell magic is invariant under composition of a state $\ket{\psi}$ with a stabilizer state $\ket{\psi_\text{STAB}}$, i.e. $\mathcal{B}(\ket{\psi}\otimes\ket{\psi_\text{STAB}})=\mathcal{B}(\ket{\psi})$. 
For additive Bell magic, we have  
\begin{equation}\label{eq:comp_add}
\mathcal{B}_\text{a}(\ket{\psi}\otimes\ket{\psi_\text{STAB}})=\mathcal{B}_\text{a}(\ket{\psi})+\mathcal{B}_\text{a}(\ket{\psi_\text{STAB}})=\mathcal{B}_\text{a}(\ket{\psi})\,,
\end{equation}
where we used the property of additivity and faithfulness. Thus, the property of composition holds for $\mathcal{B}_\text{a}$. 
Now, we apply the definition of additive Bell magic $\mathcal{B}=1-2^{-\mathcal{B}_\text{a}}$ to \eqref{eq:comp_add}, from which immediately follows that composition with a stabilizer state leaves $\mathcal{B}$ invariant as well.
}

\section{Bell magic of maximally mixed state}\label{sec:mixed}
Here, we derive the Bell magic of the maximally mixed state $\mathcal{B}(\rho_\text{m})$.
The Bell measurement applied to the maximally mixed state $\rho_\text{m}=\frac{I}{2^N}$ with identity matrix $I$ produces every possible bitstring $\boldsymbol{r}\in\{0,1\}^{2N}$ with equal probability $P(\boldsymbol{r})=4^{-N}$.
Now, we define the probability of all binary additions that yield $\boldsymbol{r}$ as $Q(\boldsymbol{r})=\sum_{\boldsymbol{q}}P(\boldsymbol{q})P(\boldsymbol{q}\oplus\boldsymbol{r})=4^{-N}$. Now, we can write the Bell magic as
\begin{equation}
\mathcal{B}=\sum_{\boldsymbol{r},\boldsymbol{q}}Q(\boldsymbol{r})Q(\boldsymbol{q})\left\lVert[\sigma_{\boldsymbol{r}},\sigma_{\boldsymbol{q}}]\right\rVert_{\infty}=4^{-2N}\sum_{\boldsymbol{r},\boldsymbol{q}}\left\lVert[\sigma_{\boldsymbol{r}},\sigma_{\boldsymbol{q}}]\right\rVert_{\infty}
\end{equation}
We split the equation into the two cases $\boldsymbol{r}=\boldsymbol{0}$ and $\boldsymbol{r}\ne{\boldsymbol{0}}$ and get
$\mathcal{B}=4^{-2N}\sum_{\boldsymbol{q}}\left\lVert[\sigma_{\boldsymbol{0}},\sigma_{\boldsymbol{q}}]\right\rVert_{\infty}+4^{-2N}\sum_{\boldsymbol{r}\ne{\boldsymbol{0}}}\sum_{\boldsymbol{q}}\left\lVert[\sigma_{\boldsymbol{r}},\sigma_{\boldsymbol{q}}]\right\rVert_{\infty}$. 
There are in total $4^N$ Pauli strings. The identity Pauli string $\sigma_{\boldsymbol{0}}=I$ commutes with every other Pauli string, i.e. $[\sigma_{\boldsymbol{0}},\sigma_{\boldsymbol{q}}]=0$. All other Pauli strings commute with half of the Pauli strings and do not commute with the other half, yielding $\sum_{\boldsymbol{q}}\left\lVert[\sigma_{\boldsymbol{r}\ne{\boldsymbol{0}}},\sigma_{\boldsymbol{q}}]\right\rVert_{\infty}=4^N$.
Thus, 
\begin{equation}
\mathcal{B}(\rho_\text{m})=4^{-2N}\sum_{\boldsymbol{r}\ne{\boldsymbol{0}}}4^N=1-4^{-N}\,.
\end{equation}

\revA{\section{Mixed Bell magic}\label{sec:mixedmagic}
We now show that $\mathcal{B}_\text{m}(\rho)=1-\frac{1-\mathcal{B}(\rho)}{\text{tr}(\rho^2)^2}$ is faithful for mixed stabilizer states of the form $\rho_\text{STAB}=U_\text{C}\ket{\psi_\text{STAB}}\bra{\psi_\text{STAB}}\otimes\rho_\text{m}U_\text{C}^\dagger$, i.e. $\mathcal{B}_\text{m}(\rho_\text{STAB})=0$. Here, $\rho_\text{m}=I_{K}2^{-K}$ is the maximally mixed state over $K$ qubits and $\ket{\psi_\text{STAB}}$ is a stabilizer state of $N-K$ qubits. 
First, we regard the case $U_\text{C}=I$. 
Then, from the invariance under composition and Appendix~\ref{sec:mixed} follows 
\begin{equation}
\mathcal{B}(\rho_\text{STAB})=\mathcal{B}(\rho_\text{m})=1-4^{-N}=1-\text{tr}(\rho_\text{m})^2\,,
\end{equation}
where in the final step we used the fact that $\text{tr}(\rho_\text{m})=2^{-K}$. Inserting above equation into the definition of mixed Bell magic, we find $\mathcal{B}_\text{m}(\rho_\text{STAB})=0$.
Faithfulness for $\rho_\text{STAB}$ with $U_\text{C}\ne I$ follows from the invariance of $\mathcal{B}$ and purity $\text{tr}(\rho^2)$ under Clifford unitaries.
}

\section{Maximal magic of pure states}\label{sec:maxmagic}
\revA{We can give an upper bound $\mathcal{B}_\text{pure}\le\mathcal{B}_\text{max}^\text{pure}$ on the Bell magic of pure states. Bell magic is the average of the commutator norm over the probability distribution $P(\boldsymbol{r})$. It becomes minimal for stabilizer states, where $P(\boldsymbol{r})$ is only non-zero for a small subset of $\boldsymbol{r}$. In contrast, it is maximal when the probability distribution is uniformly spread over as many $\boldsymbol{r}$ as possible such as for the maximally mixed state.
For pure states, the distribution has the constraint that $P(\boldsymbol{r})$ is zero for $\boldsymbol{r}$ with odd parity, i.e. $\text{tr}(\rho^2)=1-2P_\text{odd}=1$. This follows from the SWAP test~\eqref{eq:SWAP}, where odd parity outcomes are forbidden for pure states. We have $P_\text{odd}\le\frac{1}{2}(1-2^{-N})$. 
Now, we the Bell distribution $P_\text{max}(\boldsymbol{r}$ with the highest Bell magic for pure states will have zero value for odd $\boldsymbol{r}_\text{odd}$, and a constant value for even $\boldsymbol{r}_\text{even}$. Thus, by setting all even $P(\boldsymbol{r}_\text{even})=P_\text{uni}$ and all odd $P(\boldsymbol{r}_\text{odd})=0$, we find from the normalisation of probability distribution
\begin{equation}
    P_\text{uni}=\frac{2\cdot4^{-N}}{1+2^{-N}}\,.
\end{equation}
Now, following the calculation in Sec.\ref{sec:mixed}, we compute 
\begin{equation}
\mathcal{B}=\sum_{\boldsymbol{r},\boldsymbol{q}}Q(\boldsymbol{r})Q(\boldsymbol{q})\left\lVert[\sigma_{\boldsymbol{r}},\sigma_{\boldsymbol{q}}]\right\rVert_{\infty}
\end{equation}
for $Q(\boldsymbol{r})=\sum_{\boldsymbol{q}}P_\text{max}(\boldsymbol{q})P_\text{max}(\boldsymbol{q}\oplus\boldsymbol{r})$. We find for the distribution $Q(\boldsymbol{0})=P_\text{uni}$ and $Q(\boldsymbol{r}\ne\boldsymbol{0})=\frac{1-P_\text{uni}}{4^{N}-1}=\frac{1+2^{-N}-2\cdot4^{-N}}{(4^{N}-1)(1+2^{-N})}$. 
Finally, we get 
\begin{align*}
    \mathcal{B}^\text{pure}\le&\mathcal{B}_\text{max}^\text{pure}=Q(\boldsymbol{r}\ne\boldsymbol{0})^2\cdot4^{N}(4^{N}-1)=\\
    &4^{N}\frac{(1+2^{-N}-2\cdot4^{-N})^2}{(4^{N}-1)(1+2^{-N})^2}
    \,.
\end{align*}
Note that $\mathcal{B}_\text{max}^\text{pure}$ is an upper bound on the Bell magic of pure states, and in general is not saturated, as no state with the corresponding Bell measurement distribution may exist. However, we numerically find that for $N=1$ and $N=3$ the bound is saturated.

We now give explicit pure states that have the maximal amount of Bell magic which we found by using our variational quantum algorithm.
For $N=1$, this is the magic state $B_\text{a}^{(1)}=\log_2(\frac{27}{11})\approx1.29545588$ for magic state $\ket{R}$ with $\theta=\arccos(\frac{1}{\sqrt{3}})$ and $\varphi=\frac{\pi}{4}$. 
For $N=2$, the maximal states is \begin{equation}
\ket{\psi^{(2)}}=\frac{1}{2}\{1,1,1,i\}
\end{equation} with $\mathcal{B}_\text{a}^{(2)}\approx2.67807$.
For $N=3$, the state of maximal Bell magic states is the Hoggar state 
\begin{equation}
\ket{\psi^{(3)}}=\frac{1}{6}\{1+i,0,-1,1,-i,1,0,0\}
\end{equation}
with $\mathcal{B}_\text{a}^{(3)}\approx4.651794$, which is also the maximal state for the robustness of magic~\cite{howard2017application}.  
For $N=4$, we report the maximal Bell magic of $\mathcal{B}_\text{a}^{(4)}\approx6.221364$. While we did not find an exact form form this state, we here report a nearly maximal state  with a simple description
\begin{align*}
\ket{\psi^{(4)}}=&\frac{1}{8\sqrt{2}}\{4,1+i,4i,-1+i,4i,3(1+i),2i,-1-i,\\
&-1+i,4i,3(1-i),-2i,-1-i,2i,-1+i,2\}
\end{align*}
with $\mathcal{B}_\text{a}=6.221239$. 
These reported states have have substantially more Bell magic than corresponding Haar random states of same $N$ and may be useful for state preparations.

To get a better understanding what constitutes a state of high Bell magic, we write states in the form $\rho=2^{-N}(\sigma_{\boldsymbol{0}}+\sum_{\boldsymbol{r}\ne\boldsymbol{0}} \alpha_{\boldsymbol{r}} \sigma_{\boldsymbol{r}})$, where for pure states we demand $\sum_{\boldsymbol{r}}\abs{\alpha_{\boldsymbol{r}}}^2=2^N+1$. Note that for a valid state we additional demand that $\alpha_{\boldsymbol{r}}$ are chosen such that $\rho$ is positive semi-definite. Stabilizer states have exactly $2^N$ terms with $\vert\alpha_{\boldsymbol{r}}\vert=1$, and zero otherwise. In contrast, we find that states of high Bell magic have a distribution of $\alpha_{\boldsymbol{r}}$ that have similar values over all $4^N$ Pauli operators. For example for $N=1$ we have $\ket{R}\bra{R}=\frac{1}{2}(I+\frac{1}{\sqrt{3}}(\sigma_x+\sigma_y+\sigma_z))$ or for the Hoggar state for $N=3$ we have $\vert\alpha_{\boldsymbol{r}}\vert=\frac{1}{3}$.
This states also saturate the upper bound of pure state Bell magic. Note for $N=2$ the maximal pure state does not saturate the bound as it has a non-equal distribution of $\alpha_{\boldsymbol{r}}$.
}

\section{Construction of quantum circuits}\label{sec:hardwareefficient}

\begin{figure}[htbp]
	\centering	
	\subfigimg[width=0.35\textwidth]{}{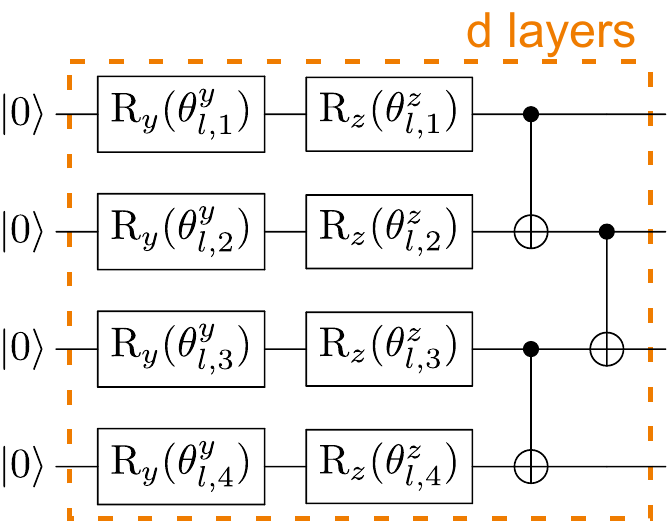} 
	\caption{Parameterized quantum circuit $\ket{\psi_\text{C}(\boldsymbol{\theta})}$ of $N$ qubits. It is composed of $d$ layers of single qubit rotations around the $y$ and $z$ axis parameterized with $\boldsymbol{\theta}$ and entangling CNOT gates arranged in a nearest-neighbor chain configuration. 
	}
	\label{fig:circuit}
\end{figure}

To prepare quantum states in an experimentally friendly way on the IonQ quantum computer, we use parameterized quantum circuits as shown Fig.\ref{fig:circuit}~\cite{haug2021capacity}. The state
\begin{equation}\label{eq:PCQ_yz}
\ket{\psi(\boldsymbol{\theta})}=\prod_{l=1}^d W [\bigotimes_{n=1}^N R_{z}(\boldsymbol{\theta}_{l,n}^z)][\bigotimes_{n=1}^N R_{y}(\boldsymbol{\theta}_{l,n}^y)]\ket{0}^{\otimes N}
\end{equation}
is generated by $d$ layers of parameterized single qubit rotations $R_\alpha(\theta)=\exp(-i\frac{\theta}{2}\sigma^\alpha)$, where $\alpha\in\{x,y,z\}$ and a set of fixed entangling gates $W$ which are CNOT gates arranged in a nearest-neighbor chain configuration. 
We can choose the $K$ parameters $\boldsymbol{\theta}$ for the parameterized quantum circuit in two fashions. First, we sample them uniformly $\boldsymbol{\theta}^{\text{rand}}\in[0,2\pi)^{K}$. This set of parameters generates highly random quantum states $\ket{\psi(\boldsymbol{\theta}^{\text{rand}})}$ that approximate Haar random states for sufficiently deep circuits~\cite{mcclean2018barren}. This circuit is used to create the highly magical states for the experimental state discrimination task on the IonQ quantum computer.

Next, to prepare stabilizer states as well as measure the transition of stabilizer into intractable quantum states on the IonQ quantum computer, we use the same circuit with different set of parameters. We choose $K-N_\text{T}$ parameters as $n\pi/2$, where $n$ is an integer, and $N_\text{T}$ parameters as multiples of $\pi/2$ shifted by $\pi/4$ with $\{n\frac{\pi}{2}+\frac{\pi}{4}\}$, yielding $\boldsymbol{\theta}^{N_\text{T}}\in\{n\frac{\pi}{2}\}^{K-N_\text{T}}\otimes\{n\frac{\pi}{2}+\frac{\pi}{4}\}^{N_\text{T}}$.
For $N_\text{T}=0$, all single qubit rotations are Clifford gates and do not introduce any magic into the circuit. The entangling gates $W$ composed of CNOTs are Clifford gates as well and thus for $N_\text{T}=0$ we get a random stabilizer states. For $N_\text{T}>0$, $N_\text{T}$ non-Clifford gates are introduced into the circuit which yield an increasing amount of magic. As one can easily check, the shift in parameter by $\pi/4$ is equivalent to adding a $T$-gate (for $z$-rotations) or a stabilizer-transformed version of the $T$-gate in the $y$-basis (for $y$-rotatations). For large $d$, our approach is equivalent to adding $N_\text{T}$ $T$-gates at random positions into a Clifford circuit sampled randomly from the Clifford group. 

For the numerical simulation of the state discrimination and magic estimation task of $N=50$ qubits, we use a modified version of Fig.\ref{fig:circuit}, where the initial state $\ket{0}$ is replaced by $N_\text{A}$ magic states. In particular, the circuit consists of an initial state of $N_\text{A}$ magic states and a Clifford circuit of depth $d$
\begin{equation}\label{eq:PCQ_yz_NT}
\ket{\psi(\boldsymbol{\theta})}=\prod_{l=1}^d W [\bigotimes_{n=1}^N R_{z}(\boldsymbol{\theta}_{l,n}^z)][\bigotimes_{n=1}^N R_{y}(\boldsymbol{\theta}_{l,n}^y)]\ket{T}^{N_\text{A}}\ket{0}^{\otimes N-N_\text{A}}
\end{equation}
Here, the position of the $N_\text{A}$ magic states $\ket{T}$ is randomly permuted within the $N$ qubits. The layered unitaries are chosen as random Clifford circuits by choosing random parameters 
$\boldsymbol{\theta}\in\{n\frac{\pi}{2}\}^{K}$ with $n$ being integer such that all parameterized gates are single qubit Clifford gates. $W$ is an entangling layer consisting of CNOT gates arranged in nearest-neighbor chain configuration. For our simulations, we use a depth of $d=4$. This circuit has a Bell magic $\mathcal{B}_\text{a}=N_\text{A}$.

\section{Error mitigation}\label{sec:error_mtg_sup}
We want to determine the Bell magic of a pure state $\ket{\psi}$ subject to depolarizing noise 
\begin{equation}\label{eq:depol_sup}
    \rho=(1-p)\ket{\psi}+pI2^{-N}
\end{equation} 
by measuring the depolarized state $\rho$. 
From the measurement of the purity $\text{tr}(\rho_\text{dp}^2)$, we can calculate 
\begin{equation}\label{eq:depolpurity_sup}
p=1-\frac{\sqrt{(2^N-1)(2^N\text{tr}(\rho_\text{dp}^2)-1)}}{2^N-1}\,.
\end{equation}
Now, we derive the error mitigation method.
First, the projector onto a Bell state can be written as
\begin{align*}
   &\Pi_{r_1r_2}=\ket{\sigma_{r_1r_2}}\bra{\sigma_{r_1r_2}}=\label{eq:projector}\numberthis\\
   &\frac{1}{4}(I\otimes I +E_{r_1r_2}^x\sigma^x\otimes\sigma^x+E_{r_1r_2}^y\sigma^y\otimes\sigma^y+E_{r_1r_2}^z\sigma^z\otimes\sigma^z) 
\end{align*}
with factors $E_{r_1r_2}^\alpha=\pm1$. 
The projector onto a product of Bell states is then given by $O_{\boldsymbol{r}}=\otimes_{n=1}^N\Pi_{r_{2n-1},r_{2n}}$.

Now, the state affected by depolarizing noise is given by $\rho=(1-p)\ket{\psi}\bra{\psi}+pI/2^N$.
The probability of measuring bitstring $\boldsymbol{r}$ via Bell measurement on the noisy state is given by \begin{align*}
P_\text{dp}&(\boldsymbol{r})=\text{Tr}(\rho\otimes\rho O_{\boldsymbol{r}})=\\
&(1-p)^2P_0(\boldsymbol{r})+p^2 4^{-N}\text{Tr}(I\otimes I O_{\boldsymbol{r}})+\\
&p(1-p)2^{-N}(\text{Tr}(\ket{\psi}\bra{\psi}\otimes I O_{\boldsymbol{r}})+\text{Tr}(I\otimes\ket{\psi}\bra{\psi} O_{\boldsymbol{r}}))
\end{align*}
with $P_0(\boldsymbol{r})=\text{Tr}(\ket{\psi}\bra{\psi}\otimes\ket{\psi}\bra{\psi}O_{\boldsymbol{r}})$. With the decomposition of the projector into Pauli strings \eqref{eq:projector}, $\text{Tr}(I\otimes \ket{\psi}\bra{\psi} \sigma^\alpha\otimes\sigma^\alpha)=0$, $\text{Tr}(I\otimes I)=4^{N}$, and $\text{Tr}(\ket{\psi}\bra{\psi}\otimes I)=2^{N}$, we get 
\begin{equation}\label{eq:prob_Bell_depol}
P_\text{dp}(\boldsymbol{r})=(1-p)^2 P_0(\boldsymbol{r})+p(2-p)4^{-N}\,.
\end{equation}
This means that depolarization occurring on one of the copies results in the same measured probability distribution as global depolarization acting on all copies. Thus, the probability of no error occurring is given by $(1-p)^2$.
Now, we define the probability 
\begin{equation}
Q_\text{dp}(\boldsymbol{r})=\sum_{\boldsymbol{q}\oplus\boldsymbol{q}'=\boldsymbol{r}}P_\text{dp}(\boldsymbol{q})P_\text{dp}(\boldsymbol{q}')
\end{equation}
of getting the binary added bitstring $\boldsymbol{r}$. The Bell magic can be then written as
\begin{equation}
\mathcal{B}_\text{dp}=\sum_{\boldsymbol{r},\boldsymbol{q}\in\{0,1\}^{2N}}Q_\text{dp}(\boldsymbol{r})Q_\text{dp}(\boldsymbol{q})\left\lVert[\sigma_{\boldsymbol{r}},\sigma_{\boldsymbol{q}}]\right\rVert_{\infty}\,.
\end{equation}
Here we used that $\sigma_{\boldsymbol{r}}=\sigma_{\boldsymbol{q}}\sigma_{\boldsymbol{q'}}=\sigma_{\boldsymbol{q}\oplus\boldsymbol{q'}}$ up to a multiplication with $\{1,-1,i,-i\}$.
Inserting $P_\text{dp}(\boldsymbol{r})$ with the depolarizing noise, we get
$Q_\text{dp}(\boldsymbol{r})=(1-p)^4Q_0(\boldsymbol{r})+(1-(1-p)^4)4^{-N}$, where $Q_0(\boldsymbol{r})=\sum_{\boldsymbol{q}\oplus\boldsymbol{q}'}P_0(\boldsymbol{q})P_0(\boldsymbol{q})$ is the probability for the pure state. Here, we used the fact that $\sum_{\boldsymbol{r}\in\{0,1\}^{2N}}P_0(\boldsymbol{r})=1$ and $\sum_{\boldsymbol{r}\in\{0,1\}^{2N}}=4^N$. We now define 
\begin{equation}
    p_\text{c}=1-(1-p)^4
\end{equation} 
to simplify to $Q_\text{dp}(\boldsymbol{r})=p_\text{c}Q_0(\boldsymbol{r})+(1-p_\text{c})4^{-N}$.
We can now write out the Bell magic in terms of the probability $Q_0$ for the noise-free state
\begin{align*}
\mathcal{B}_\text{dp}&=\sum_{\boldsymbol{r},\boldsymbol{q}\in\{0,1\}^{2N}}Q_\text{dp}(\boldsymbol{r})Q_\text{dp}(\boldsymbol{q})\left\lVert[\sigma_{\boldsymbol{r}},\sigma_{\boldsymbol{q}}]\right\rVert_{\infty}=\\
&\sum_{\boldsymbol{r},\boldsymbol{q}}\left\lVert[\sigma_{\boldsymbol{r}},\sigma_{\boldsymbol{q}}]\right\rVert_{\infty}(p_\text{c}^2Q_0(\boldsymbol{r})Q_0(\boldsymbol{q})+\\
&(1-p_\text{c})^2 4^{-2N}+2p_\text{c}(1-p_\text{c})4^{-N}Q_0(\boldsymbol{r})=\\
&(1-p_\text{c})^2\mathcal{B}^\text{mtg}+p_\text{c}^2\mathcal{B}(\rho_\text{m})+2p_\text{c}(1-p_\text{c})\mathcal{B}^{\text{R}}\,.
\end{align*}
Here, $\mathcal{B}(\rho_\text{m})=\sum_{\boldsymbol{r},\boldsymbol{q}}4^{-2N}\left\lVert[\sigma_{\boldsymbol{r}},\sigma_{\boldsymbol{q}}]\right\rVert_{\infty}=1-4^{-N}$ is the Bell magic of the maximally mixed state, $\mathcal{B}^\text{mtg}=\sum_{\boldsymbol{r},\boldsymbol{q}}Q_0(\boldsymbol{r})Q_0(\boldsymbol{q})\left\lVert[\sigma_{\boldsymbol{r}},\sigma_{\boldsymbol{q}}]\right\rVert_{\infty}$ is the mitigated Bell magic of the noise-free state and we defined 
\begin{equation}
\mathcal{B}^\text{R}=4^{-N}\sum_{\boldsymbol{r},\boldsymbol{q}}Q_0(\boldsymbol{r})\left\lVert[\sigma_{\boldsymbol{r}},\sigma_{\boldsymbol{q}}]\right\rVert_{\infty}\,.
\end{equation}
We now show how to calculate $\mathcal{B}^\text{R}$. First, we split this term into the case $\boldsymbol{r}={\boldsymbol{0}}$ and $\boldsymbol{r}\ne{\boldsymbol{0}}$
\begin{align*}
\mathcal{B}^\text{R}=&4^{-N}\sum_{\boldsymbol{q}}Q_0({\boldsymbol{0}})\left\lVert[\sigma_{\boldsymbol{0}},\sigma_{\boldsymbol{q}}]\right\rVert_{\infty}+\\
&4^{-N}\sum_{\boldsymbol{r}\ne {\boldsymbol{0}}}Q_0(\boldsymbol{r})\sum_{\boldsymbol{q}}\left\lVert[\sigma_{\boldsymbol{r}},\sigma_{\boldsymbol{q}}]\right\rVert_{\infty}
\end{align*}
A given Pauli string $\sigma_{\boldsymbol{r}}$ with $\boldsymbol{r}\ne\boldsymbol{0}$ commutes with half of all $4^N$ Pauli string, while non-commutes with the other half. Thus, $\sum_{\boldsymbol{q}}\left\lVert[\sigma_{\boldsymbol{r}\ne{\boldsymbol{0}}},\sigma_{\boldsymbol{q}}]\right\rVert_{\infty}=4^N$.
For the case $\boldsymbol{r}=\boldsymbol{0}$ the Pauli string $\sigma_{\boldsymbol{0}}=I$ is the identity and thus always commutes $\left\lVert[\sigma_{\boldsymbol{0}},\sigma_{\boldsymbol{q}}]\right\rVert_{\infty}=0$. Using $4^{-N}\sum_{\boldsymbol{r}\ne{\boldsymbol{0}}}1=1-Q_0({\boldsymbol{0}})$, we get 
\begin{equation}
  \mathcal{B}^\text{R}=1-Q_0({\boldsymbol{0}})=1-\sum_{\boldsymbol{q}\oplus\boldsymbol{q}'={\boldsymbol{0}}}P_0(\boldsymbol{q})P_0(\boldsymbol{q}')=1-\sum_{\boldsymbol{q}}P_0(\boldsymbol{q})^2\,.  
\end{equation}
For a stabilizer state $\sum_{\boldsymbol{q}}P_0(\boldsymbol{q})^2=2^{-N}$ as $P(\boldsymbol{q})=2^{-N}\vert\bra{\psi}\sigma_{\boldsymbol{r}}\ket{\psi^*}\vert^2$ and there are $2^N$ Pauli strings with non-zero expectation values. On the other hand, for a maximally mixed state we find $\sum_{\boldsymbol{q}}P_0(\boldsymbol{q})^2=4^{-N}$ as every bitstring $\boldsymbol{q}$ appears with equal probability. Thus, we can bound $1-2^{-N}\le\mathcal{B}^\text{R}\le 1-4^{-N}$.
Now we want to calculate $\sum_{\boldsymbol{q}}P_0(\boldsymbol{q})^2$ by measuring the depolarized state. The sum of the squares of the probabilities of the bitstrings is given by
$\sum_{\boldsymbol{q}}P_\text{dp}(\boldsymbol{q})^2$. Inserting \eqref{eq:prob_Bell_depol}, we get
$\sum_{\boldsymbol{q}}P_\text{dp}(\boldsymbol{q})^2=(1-p_\text{c})\sum_{\boldsymbol{q}}P_0(\boldsymbol{q})^2+p_\text{c}4^{-N}$. 
We invert this equation to get 
\begin{equation}
\sum_{\boldsymbol{q}}P_0(\boldsymbol{q})^2=\frac{\sum_{\boldsymbol{q}}P_\text{dp}(\boldsymbol{q})^2-4^{-N}p_\text{c}}{1-p_\text{c}}
\end{equation}
Putting all our results together, the mitigated Bell magic is given by
\begin{equation}\label{eq:mtg}
\mathcal{B}^\text{mtg}=\frac{1}{(1-p_\text{c})^2}(\mathcal{B}^\text{dp}-p_\text{c}^2\mathcal{B}(\rho_\text{m})-2p_\text{c}(1-p_\text{c})\mathcal{B}^\text{R})\,,
\end{equation}
where $\mathcal{B}^\text{R}=1-(1-p_\text{c})^{-1}(\sum_{\boldsymbol{q}}P_\text{dp}(\boldsymbol{q})^2-4^{-N}p_\text{c})$. For a large number of qubits $N$, the sum of probabilities becomes exponentially small $4^{-N}\le \sum_{\boldsymbol{q}}P(\boldsymbol{q})^2\le 2^{-N}$ and becomes challenging to measure. In this limit we can approximate $\mathcal{B}^\text{R}\approx \mathcal{B}(\rho_\text{m})\approx1$ and finally get
\begin{equation}
\mathcal{B}^\text{mtg}\approx\frac{\mathcal{B}^\text{dp}-p_\text{c}(2-p_\text{c})}{(1-p_\text{c})^2}\,.
\end{equation}

\section{Supervised learning for decision boundaries}\label{sec:suplearn}

We want to learn to classify unknown states using Bell magic measured on (noisy) states with a finite number of measurement samples $N_\text{Q}$. We have two classes of states with different amounts of Bell magic, i.e. class $\beta$ with stabilizer states with low Bell magic  and class $\alpha$ with random states with high Bell magic. To train the classifier, we are given a training set of $N_\text{train}$ states where we know to which class the states belong with label $y_i\in\{-1,1\}$. The label $y_i=-1$ indicates class $\beta$ and  $y_i=1$ class $\alpha$. We now measure Bell magic using $N_\text{Q}$ measurement samples for each state of the training set and estimate the Bell magic for each state $\hat{\mathcal{B}}^{(i)}$. 
Now, we want to find the best threshold $\mathcal{B}^*$ that separates the two classes such that $\hat{\mathcal{B}}^{(i)}\le \mathcal{B}^*$ is correctly assigned $\hat{y}_i=-1$ and $\hat{\mathcal{B}}_2^{(i)}>\mathcal{B}^*$ as $\hat{y}_i=1$.
To find the best threshold, we maximize 
\begin{equation}
    \mathcal{B}^*_\text{opt}=\text{max}_{\mathcal{B}^*} \sum_{i=1}^{N_\text{train}}\text{sign}(\hat{\mathcal{B}}^{(i)}-\mathcal{B}^*)y_i\,.
\end{equation}
To evaluate the performance of the classifier, we test the threshold $\mathcal{B}^*_\text{opt}$ on an unlabeled test dataset of $N_\text{test}$ states which have not been used during training. We define $P_\text{error}=N^\text{wrong}_\text{test}/N_\text{test}$ as the probability of wrongly classifying $N^\text{wrong}_\text{test}$ states. The trivial strategy of randomly guessing would achieve an error probability of $P_\text{error}=\frac{1}{2}$.

\section{Data for learning state discrimination}\label{sec:learndata}
\begin{figure}[htbp]
	\centering	
	\subfigimg[width=0.33\textwidth]{}{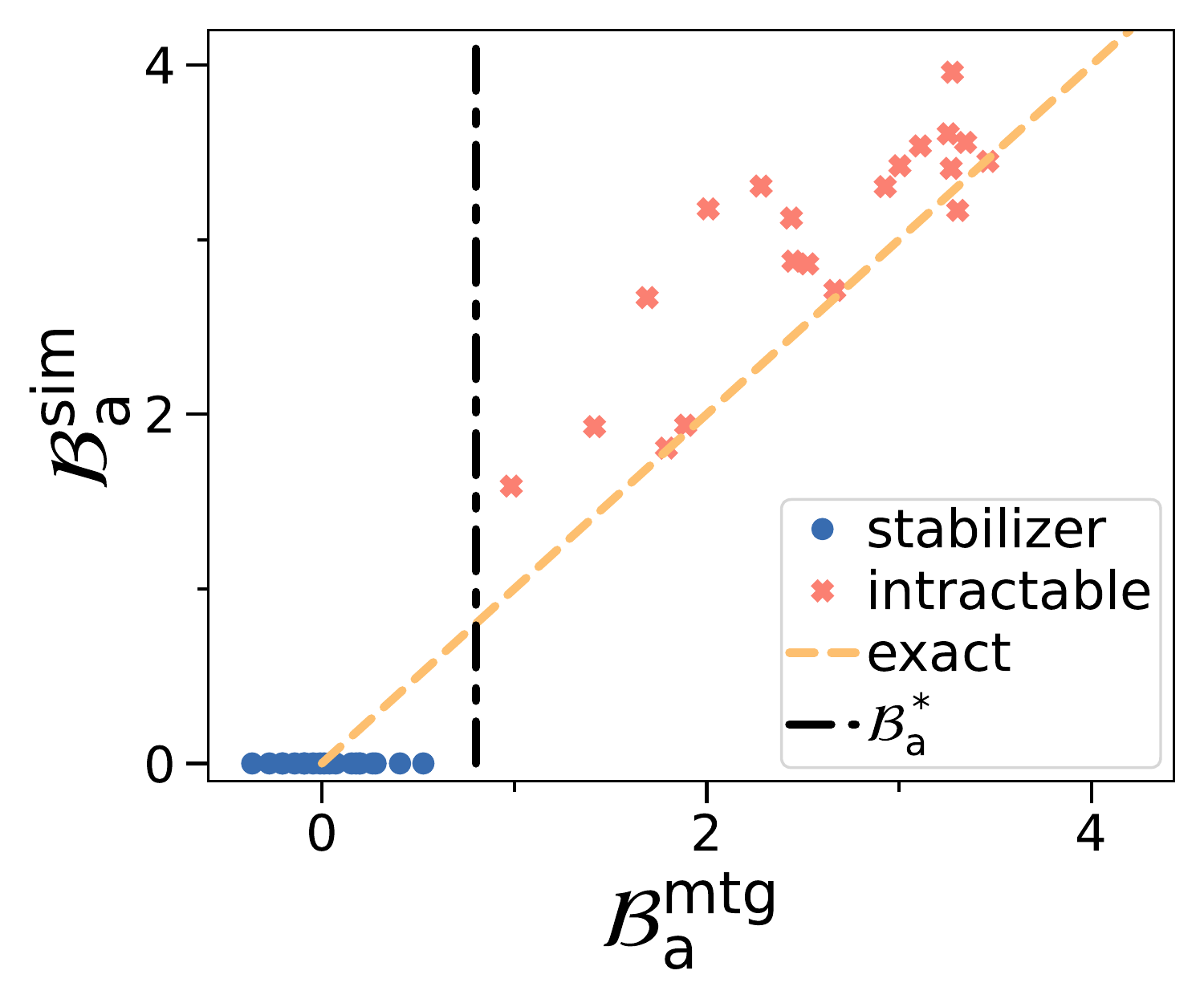}
	\caption{Simulated $\mathcal{B}_\text{a}^\text{sim}$ plotted against experimentally measured mitigated additive Bell magic $\mathcal{B}_\text{a}^\text{mtg}$. These datapoints were used for the state discrimination task in the main text. We show the result for $N_\text{Q}=1000$ measurements and $N=3$ qubits.
	}
	\label{fig:learndata}
\end{figure}

In Fig.\ref{fig:learndata} we show the data we used for our experimental demonstration of the state discrimination protocol on the IonQ quantum computer. We show the simulated $\mathcal{B}_\text{a}^\text{sim}$ and mitigated additive Bell magic $\mathcal{B}_\text{a}^\text{mtg}$. Blue is the stabilizer data, while orange are highly magical states generated by a hardware efficient circuit with random parameters as defined in Appendix~\ref{sec:hardwareefficient}. The linear curve is the relationship expected for perfect results. We observe that the Bell magic is slightly underestimated in experiment. The vertical dashed line is the optimal threshold $\mathcal{B}^*_\text{a}$ to discriminate stabilizer and non-stabilizer states from the experiment for $N_\text{Q}=1000$. For smaller $N_\text{Q}$, the measurement error increases and no $\mathcal{B}^*_\text{a}$ to perfectly distinguish the two classes can be found, yielding a finite classification error.

\section{Shift rule for Bell magic}\label{sec:shiftrule}
We now derive the shift-rule for Bell measurements and the Bell magic.
The shift-rule provides exact gradients when the circuit is composed of parameterized Pauli rotations~\cite{schuld2019evaluating}. For standard measurements on single quantum states, the shift rule is given by $\partial_k\langle C(\boldsymbol{\theta})\rangle=v(\langle C(\boldsymbol{\theta}+\boldsymbol{e}_k\frac{\pi}{4v})\rangle-\langle C(\boldsymbol{\theta}-\boldsymbol{e}_k\frac{\pi}{4v})\rangle)$, where $\boldsymbol{e}_k$ is the $k$th unit vector and $r$ is an arbitrary number~\cite{mitarai2018quantum}.

We can write for any operators $U$, $V$ and $O$ \cite{schuld2019evaluating} \begin{align*}\label{eq:reorder}
\bra{\psi}U^\dagger O V\ket{\psi}+\text{h.c.}&=\frac{1}{2}[\bra{\psi}(U+V)^\dagger O (U+V)\ket{\psi}-\\
&\bra{\psi}(U-V)^\dagger O(U-V)\ket{\psi}]\,,\numberthis
\end{align*}
where $\text{h.c.}$ indicates the hermitian conjugate of the preceding terms.

For Bell magic, we have to estimate the probability $P(\boldsymbol{r})=\bra{\psi}\bra{\psi}O_{\boldsymbol{r}}\ket{\psi}\ket{\psi}$ of measuring a Bell state $\ket{\sigma_{\boldsymbol{r}}}$, where $O_{\boldsymbol{r}}=\ket{\sigma_{\boldsymbol{r}}}\bra{\sigma_{\boldsymbol{r}}}$ is the projector onto the Bell state.
For parameterized quantum circuits with $d$ layers of the form $\ket{\psi(\boldsymbol{\theta})}=\prod_{n=1}^{d} V_n(\boldsymbol{\theta}_n)W_n\ket{0}$ with entangling gates $W_n$, parameters $\boldsymbol{\theta}$ and parameterized rotations $V_n(\boldsymbol{\theta}_n)=e^{-i\frac{\boldsymbol{\theta}_k}{2}\sigma_n}$ generated by some Pauli string $\sigma_n$. 
The derivative on the quantum state $\ket{\psi}$ can be written as
\begin{align*}
\partial_k\ket{\psi(\boldsymbol{\theta})}=&\prod_{n=k+1}^{d}[ V_n(\boldsymbol{\theta}_n)W_n](-i\frac{1}{2}\sigma_k)\prod_{n=1}^{k}[ V_n(\boldsymbol{\theta}_n)W_n]\ket{0}\\
&\equiv U_k(-i\frac{1}{2}\sigma_k)\ket{\phi_k}\,,
\end{align*}
where in the last step we define $\ket{\phi_k}=\prod_{n=1}^{k} V_n(\boldsymbol{\theta}_n)W_n\ket{0}$ and $U_k=\prod_{n=k+1}^{d} V_n(\boldsymbol{\theta}_n)W_n$.
Now, the derivative of $P(\boldsymbol{r})$ using the product rule is given by
\begin{align*}
\partial_k P(\boldsymbol{r})=&2\bra{\phi_k}\bra{\phi_k} (U_k \otimes U_k)^\dagger O_{\boldsymbol{r}} (U_k \otimes U_k)\\
&[(-i\frac{1}{2}\sigma_k)\otimes I] \ket{\phi_k} \ket{\phi_k} + \text{h.c.} \,.
\end{align*}
We now define for simplicity $O_{\boldsymbol{r}}'=(U_k \otimes U_k)^\dagger O_{\boldsymbol{r}} (U_k \otimes U_k)$, introduce an arbitrary factor $v>0$  and apply \eqref{eq:reorder}
\begin{align*}
&\partial_k P(\boldsymbol{r})=2v\bra{\phi_k}\bra{\phi_k} O_{\boldsymbol{r}}'[(-i\frac{1}{2v}\sigma_k)\otimes I] \ket{\phi_k} \ket{\phi_k} + \text{h.c.}=\\
&v\bra{\phi_k}\bra{\phi_k}[(I-i\frac{1}{2v}\sigma_k)^\dagger\otimes I] O_{\boldsymbol{r}}'[(I-i\frac{1}{2v}\sigma_k)\otimes I] \ket{\phi_k} \ket{\phi_k}-\\
&v\bra{\phi_k}\bra{\phi_k}[(I+i\frac{1}{2v}\sigma_k)^\dagger\otimes I] O_{\boldsymbol{r}}'[(I+i\frac{1}{2v}\sigma_k)\otimes I] \ket{\phi_k} \ket{\phi_k}\,.
\end{align*}
For any Pauli strings $\sigma$, we can rewrite the generators into a unitary as follows~\cite{schuld2019evaluating}
\begin{equation}
    e^{-i\frac{\pi}{4v}\frac{1}{2}\sigma}=\frac{1}{\sqrt{2}}(I-i\frac{1}{2v}\sigma)\,.
\end{equation}
We now find
\begin{align*}
&\partial_k P(\boldsymbol{r})=\numberthis\label{eq:shift_rule_sup}\\
&2v\bra{\phi_k}\bra{\phi_k}[e^{-i\frac{\pi}{4v}\frac{1}{2}\sigma_n}\otimes I]^\dagger O_{\boldsymbol{r}}'[e^{-i\frac{\pi}{4v}\frac{1}{2}\sigma_n}\otimes I] \ket{\phi_k} \ket{\phi_k}-\\
&2v\bra{\phi_k}\bra{\phi_k}[e^{i\frac{\pi}{4v}\frac{1}{2}\sigma_n}\otimes I]^\dagger O_{\boldsymbol{r}}'[e^{i\frac{\pi}{4v}\frac{1}{2}\sigma_n}\otimes I] \ket{\phi_k} \ket{\phi_k}=\\
&2v\bra{\psi(\boldsymbol{\theta}+\frac{\pi}{4v}\boldsymbol{e}_k)}\bra{\psi(\boldsymbol{\theta})} O_{\boldsymbol{r}}\ket{\psi(\boldsymbol{\theta}+\frac{\pi}{4v}\boldsymbol{e}_k)}\ket{\psi(\boldsymbol{\theta})}-\\
&2v\bra{\psi(\boldsymbol{\theta}-\frac{\pi}{4v}\boldsymbol{e}_k)}\bra{\psi(\boldsymbol{\theta})} O_{\boldsymbol{r}}\ket{\psi(\boldsymbol{\theta}-\frac{\pi}{4v}\boldsymbol{e}_k)}\ket{\psi(\boldsymbol{\theta})}\,,
\end{align*}
where in the last step we introduced the $k$th unit vector $\boldsymbol{e}_k$ and we absorbed $e^{-i\frac{\pi}{4v}\frac{1}{2}\sigma_n}$ into the definition of the $k$th parameterized rotation, i.e. $U_ke^{-i\frac{\pi}{4v}\frac{1}{2}\sigma_n}\ket{\phi_k}=\ket{\psi(\boldsymbol{\theta}+\frac{\pi}{4v}\boldsymbol{e}_k)}$.
Finally, with the product rule the gradient of Bell magic is given by
\begin{equation}\label{eq:BellMagicGrad_sup}
\partial_k \mathcal{B}=4\sum_{\substack{\boldsymbol{r},\boldsymbol{r'},\boldsymbol{q},\boldsymbol{q'}\\\in\{0,1\}^{2N}}}[\partial_k P(\boldsymbol{r})]P(\boldsymbol{r'})P(\boldsymbol{q})P(\boldsymbol{q'})\left\lVert[\sigma_{\boldsymbol{r}}\sigma_{\boldsymbol{r'}},\sigma_{\boldsymbol{q}}\sigma_{\boldsymbol{q'}}]\right\rVert_{\infty}
\end{equation}

\section{Trainability of Bell magic}\label{sec:express}
Here, we present the calculation of the gradient of the Bell magic for the ansatz $\ket{\psi(\theta,U_\text{C}}=U_\text{C}\exp(-i\frac{1}{2}\theta \sigma^y_1)\ket{0}^{\otimes N}$.
A straightforward calculation yields $\mathcal{B}(\theta,U_{\text{C}})=\frac{1}{2}\sin(2\theta)^2$. Here, we use that Bell magic is invariant under the choice of Clifford circuit $U_\text{C}$.
For the gradient, we find $\partial_\theta\mathcal{B}(\theta,U_{\text{C}})=\sin(4\theta)$. 
We calculate the variance of the gradient by integrating over $\theta$, which yields
\begin{align*}
&\text{Var}(\partial_\theta\mathcal{B}(\theta,U_{\text{C}}))_{\theta, U_\text{C}}=\\
&\frac{1}{2\pi}(\int_{0}^{2\pi}[\partial_\theta\mathcal{B}(\theta,U_{\text{C}})]^2\text{d}\theta-(\int_{0}^{2\pi}\partial_\theta\mathcal{B}(\theta,U_{\text{C}})\text{d}\theta)^2)=\frac{1}{2}\,.
\end{align*}

\section{State discrimination for a single magic state}\label{sec:singleqdiscr}
Here we derive the error probability of classifying a Clifford circuit with exactly one magical state as input.
The state $\ket{\psi_C(\phi)}=U_\text{C}\ket{\phi}\otimes\ket{0}^{N-1}$ consists of an arbitrary Clifford circuit $U_\text{C}$ and an initial state $\ket{A_\phi}\otimes\ket{0}^{N-1}$ with exactly one non-stabilizer qubit $\ket{A_\phi}=\cos(\frac{\phi}{2})\ket{0}+\sin(\frac{\phi}{2})\ket{1}$. $\phi$ controls the amount of Bell magic introduced into the circuit. In particular for $\phi=n\pi/2$, $n$ being an integer, no magic is introduced, whereas for $\phi=\pi/4$ the Bell magic introduced in the circuit is equivalent to the $\ket{T}$ state.

We now derive the error for the case $U_\text{C}=I$ and $N=1$ with $\ket{\psi_C(\phi)}=\ket{A_\phi}$. General $N$ and $U_\text{C}$ follows from faithfulness, composition and invariance under transformation with $U_C$ for the Bell magic.
First, we apply the Bell transformation and get
\begin{equation}
U_\text{Bell}\ket{A_\phi}\otimes\ket{A_\phi}=\frac{1}{\sqrt{2}}(\ket{00}-\cos(\phi)\ket{10}+\sin(\phi)\ket{01})
\end{equation}
The probability of measuring respective bitstrings are $p_{00}=\frac{1}{2}$, $p_{10}=\frac{1}{2}\cos(\phi)^2$ and $p_{10}=\frac{1}{2}\sin(\phi)^2$.
Now, we sample this state $N_\text{Q}$ times and apply the algorithm for the Bell magic. We assume the case of large number of resampling steps $N_\text{R}$ such that the algorithm adds all possible combination pairs of bitstrings together, then checks whether any of those pairs correspond to non-commuting Pauli strings. 
One can easily check that such a non-commuting pair of Pauli strings is only found when one samples each possible bitstring $\{00\}$, $\{01\}$ and $\{10\}$ at least once. Thus, the error probability  of finding only commuting Pauli strings and thus wrongly estimating $\mathcal{B}=0$, is given by
\begin{align*}
P_\text{E}(N_\text{Q})=&\sum_{k=0}^{N_\text{Q}}{N_\text{Q} \choose k}[p_{00}^{N-k}p_{01}^k+p_{00}^{N-k}p_{10}^k+p_{10}^{N-k}p_{01}^k]-\\
&p_{00}^N-p_{10}^N-p_{01}^N\,.
\end{align*}
Here, we sum over all possible combinations of measuring only two kinds of bitstrings. The last three terms are subtracted as these terms appear twice in the sums.
After inserting the probabilities and simplifying, we get
\begin{align*}
P_\text{E}(\phi)=& 4^{-N_\text{Q}}[(3-\cos(2\phi))^{N_\text{Q}}+(3+\cos(2\phi))^{N_\text{Q}}]-\\
&2^{-N_\text{Q}}[\sin(\phi)^{2N_\text{Q}}+\cos(2\phi)^{2N_\text{Q}}]
\end{align*}

\begin{figure*}[htbp]
	\centering	
	\subfigimg[width=0.38\textwidth]{}{BellMagicIonQ.pdf}
	\subfigimg[width=0.38\textwidth]{}{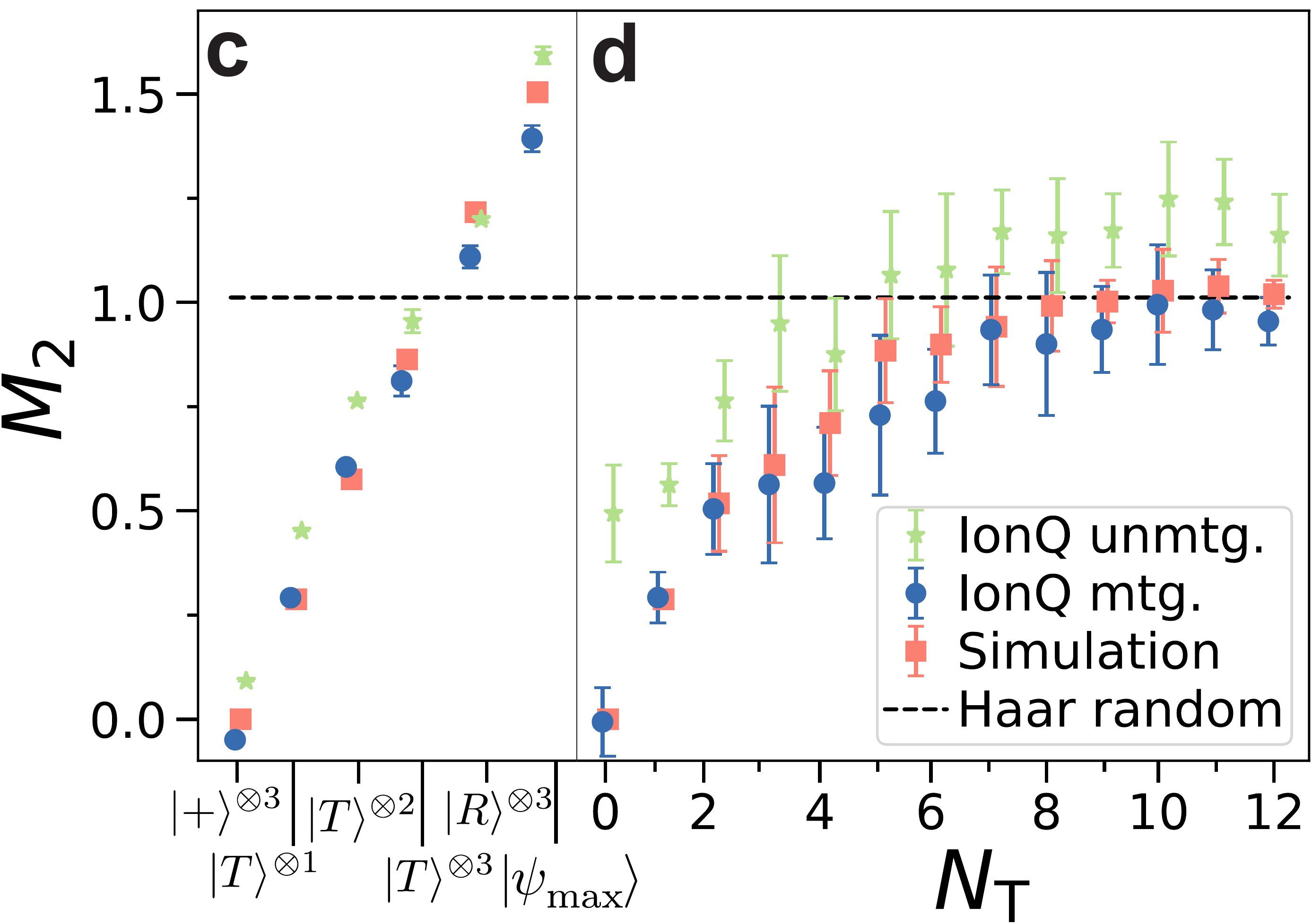}
	\caption{Experiment to measure \idg{a,b} additive Bell magic $\mathcal{B}_\text{a}$ and \idg{c,d} stabilizer $2$-Rényi entropy $M_2$ on the IonQ quantum computer for various types of states. \idg{a} In the left part of the graph, we show product states of stabilizer states $\ket{+}^{\otimes N}$, magic states $\ket{T}^{\otimes N}$, $\ket{R}^{\otimes N}$ as well as the state of maximal Bell magic $\ket{\psi_\text{max}}$ for $N=3$. \idg{b} In the right part of the graph, we show magic as a function of $N_\text{T}$ $T$-gates inserted at random positions in a Clifford circuit.  
	For all measures of magic, we show the unmitigated and mitigated magic from IonQ quantum computer as well as an exact simulation of the quantum states. The mean value and the standard deviation of is taken over 6 random instances of the state for $N=3$ qubits. The dashed line is the Bell magic averaged over Haar random states. The experiment is performed with $N_\text{Q}=10^3$ measurement samples and no further error or readout error mitigation. The purity measured on the IonQ quantum computers gives us a depolarization error of $p\approx0.1$.
	}
	\label{fig:ionqmeas_sup}
\end{figure*}

\section{Stabilizer entropy and Bell measurement}\label{sec:connect_stabilizer_sup}
Stabilizer entropy is a class of measures of magic that have been recently introduced~\cite{leone2021renyi}. They can be measured using a randomized measurement approach.
As alternative approach, we show how to measure the stabilizer entropy with Bell measurements.
The stabilizer $2$-Rényi entropy is given by
\begin{equation}
M_2=-\log(2^{N}\sum_{\boldsymbol{r}}(2^{-N}\bra{\psi}\sigma_{\boldsymbol{r}}\ket{\psi}^2)^2)
\end{equation} 
and the linear stabilizer entropy by 
\begin{equation}M_\text{lin}=1-2^{N}\sum_{\boldsymbol{r}}(2^{-N}\bra{\psi}\sigma_{\boldsymbol{r}}\ket{\psi}^2)^2\,.
\end{equation}
With \eqref{eq:to_show} and setting $\boldsymbol{n}=\boldsymbol{0}$, we can write the stabilizer $2$-Rényi entropy with the outcome probability $P(\boldsymbol{r})$ of Bell measurements as 
\begin{equation}
M_2=-\log(2^{N}\sum_{\boldsymbol{r}}P(\boldsymbol{r})^2)\,.
\end{equation}
and the linear stabilizer entropy
\begin{equation}
M_\text{lin}=1-2^{N}\sum_{\boldsymbol{r}}P(\boldsymbol{r})^2
\end{equation}
Note that the explicit estimation of $P(\boldsymbol{r})$ requires a number of measurement samples that scales exponentially with number of qubits $N$. Using our shift-rule~\eqref{eq:shift_rule_sup}, one could also maximize $M_2$ in a variational quantum algorithm.

We can also mitigate errors of $M_2$ on noisy quantum computers. Assuming depolarizing noise $p$ as outlined in Sec.\ref{sec:error_mtg_sup}, we measure the probabilities $P_\text{dp}(\boldsymbol{r})$ of the noisy state. Using the purity estimated from the SWAP test of the Bell measurements~\eqref{eq:depolpurity_sup} we can estimate $p$. Then, the mitigated probabilities $P_0(\boldsymbol{r})$ can be computed via \eqref{eq:prob_Bell_depol} as
\begin{equation}
P_0(\boldsymbol{r})=\frac{P_\text{dp}(\boldsymbol{r})-p(2-p)4^{-N}}{(1-p)^2}\,.
\end{equation}
Note that negative probabilities can appear due to shot noise or the noise not being perfectly depolarizing. In this case, we set all $P_0(\boldsymbol{r})<0$ to zero.

In Fig.\ref{fig:ionqmeas_sup}, we compare the experimental results for Bell measurements on the IonQ quantum computer for stabilizer entropy and Bell magic. For all measures, we use the same experimental data.
We show additive Bell magic $\mathcal{B}_\text{a}$ in Fig.\ref{fig:ionqmeas_sup}a,b and stabilizer $2$-Rényi entropy $M_2$ in Fig.\ref{fig:ionqmeas_sup}c,d. We find that error mitigation improves the result for all measures. For the exact simulation, all measures produce similar behavior, in particular we observe that both $\mathcal{B}_\text{a}$ and $M_2$ is additive for the various $\ket{T}$ states with $N$. Both measures share the same state of maximal Bell magic. Further, we see that with increasing $N_\text{T}$ the average magic of both measures converge to their respective value found for Haar random states.

\revA{\section{Entanglement and Bell measurements}\label{sec:entangle}
\begin{figure*}[htbp]
	\centering	
	\subfigimg[width=0.38\textwidth]{}{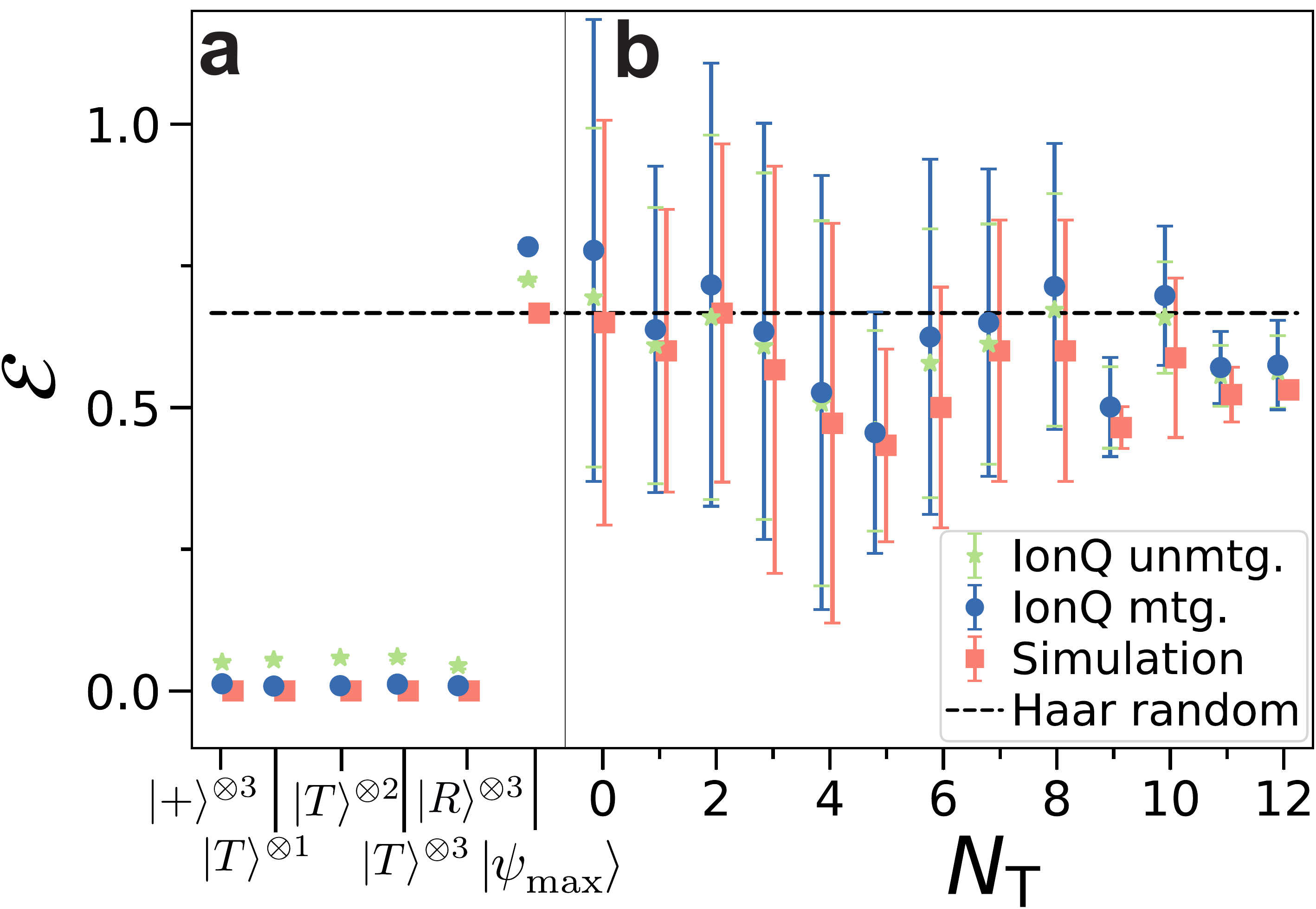}
	\caption{Measurement of the Wallach-Meyer measure $\mathcal{E}$ for entanglement on the IonQ quantum computer. We compute $\mathcal{E}$ for various type of states. We use the same states and parameters as in Fig.\ref{fig:ionqmeas_sup}. The dashed line is $\mathcal{E}$ for Haar random states.
	}
	\label{fig:ionqmeas_ent_sup}
\end{figure*}
We now investigate the entanglement of the states we studied in the main text for Bell magic.
The Meyer-Wallach measure $\mathcal{E}$~\cite{meyer2002global} has been proposed as a measure to characterize the entanglement of states prepared on quantum computers~\cite{sim2019expressibility}. 
First, one defines a mapping $\iota_j(e)$ that acts on the computational basis states $
\iota_j(b) \ket{b_1 \cdots b_n} =  \delta_{bb_j} \ket{b_1 \cdots \tilde{b}_j \cdots b_n}$,
where $b_j \in \{0, 1\}$ and $\tilde{b}_{j}$ denotes the absence of the $j$-th qubit. The Meyer-Wallach measure is then defined as
\begin{equation}
\mathcal{E}(\ket{\psi}) \equiv \frac{4}{n} \sum_{j=1}^n D \big( \iota_j(0) \ket{\psi}, \iota_j(1) \ket{\psi} \big),
\end{equation}
where $D$ is the generalized distance of the coefficients of two states $\ket{u} = \sum u_i \ket{e_{i}}$ and $\ket{v} = \sum v_i \ket{e_{i}}$,
\begin{equation} \label{eq:meyer_wallach_distance}
D(\ket{u}, \ket{v}) = \frac{1}{2} \sum_{i,j} \vert u_i v_j - u_j v_i \vert^2.
\end{equation}
It can be rewritten as~\cite{brennen2003observable}
\begin{equation}
    \mathcal{E}(\psi)=2(1-\frac{1}{N}\sum_{i=k}^N\text{tr}(\rho_k^2))
\end{equation}
where $\rho_k=\text{tr}_{k}(\rho)$ is the partial trace of $\ket{\psi}$ over all qubits except the $k$th qubit. It is zero for pure product states, and can maximally reach $\mathcal{E}=1$ for classes of highly entangled states such as the GHZ state.
For Haar random states and random stabilizer states, we have $\text{tr}(\rho_k^2)=\frac{2^{N-1}+2}{2^N+1}$ and thus $\mathcal{E}(\ket{\psi_\text{Haar}})=\frac{2^N-2}{2^N+1}$~\cite{nechita2007asymptotics}. 
We now assume that a pure state $\ket{\psi}$ is subject to depolarizing error $p$, resulting in the noisy state $\rho_\text{dp}$ as defined in~\eqref{eq:depol_sup}. By measuring the Meyer-Wallach measure $\mathcal{E}_\text{dp}$ on $\rho_\text{dp}$, we want to compute the mitigated measure $\mathcal{E}_\text{mtg}$ for the corresponding pure state $\ket{\psi}$. We now apply the error mitigation method in Appendix.~\ref{sec:error_mtg_sup} with
\begin{equation}
\rho_{k,\text{dp}}=\rho_k(1-p)+\frac{1}{2}I_kp
\end{equation}
After squaring and taking the trace over $\rho_{k,\text{dp}}$, we find
\begin{equation}
\mathcal{E}_\text{mtg}=\frac{\mathcal{E}_\text{dp} -(2-p)\frac{p}{2}}{(1-p)^2}\,.
\end{equation}
Note that the $\text{tr}(\rho_k^2)$ can be efficiently computed via Bell measurements~\cite{garcia2013swap}. In particular, $\text{tr}(\rho_k^2)=1-2P_\text{odd,k}$, where $P_\text{odd,k}$ is the probability of odd parity of the outcomes measured on the $k$th qubits of the two copies. 

We experimentally measure $\mathcal{E}$ on the IonQ quantum computer in Fig.\ref{fig:ionqmeas_ent_sup}. We use the same states and parameters as used for computing Bell magic in Fig.\ref{fig:ionqmeas_sup}.
As expected, we find experimentally that product states have $\mathcal{E}\approx0$. In contrast, entangled states such as $\ket{\psi}_\text{max}$ and the random Clifford states with $T$-gates have high $\mathcal{E}$. We find that $\mathcal{E}$ is nearly independent of $N_\text{T}$, and close to the average value expected for Haar random states. We observe some variance in our result as we only consider states with short circuit depth and measured only a low number of states. 
Our results highlight the complementary properties of Bell magic and entanglement of different types of states, which can be easily measured with Bell measurements on noisy quantum computers.
}

\end{document}